\documentclass{article}

\usepackage{arxiv}

\usepackage[utf8]{inputenc} 
\usepackage[T1]{fontenc}    
\usepackage{hyperref}       
\usepackage{url}            
\usepackage{booktabs}       
\usepackage{amsfonts}       
\usepackage{nicefrac}       
\usepackage{microtype}      
\usepackage{lipsum}

\usepackage{epsfig}
\usepackage[displaymath]{lineno}
\usepackage{hyperref}
\usepackage{graphicx,pstricks}
\usepackage{graphics}
\usepackage{amssymb}
\usepackage{amsmath, amsthm}
\usepackage{amsfonts}
\usepackage{algorithm}
\usepackage{booktabs,tabu,multirow}
\usepackage{subfig}
\usepackage{caption}
\usepackage{colortbl}
\usepackage{setspace}

\title{The effects of leaflet material properties on the simulated function of regurgitant mitral valves}

\author{
Wensi Wu\\
Department of Anesthesiology and\\
Critical Care Medicine,\\
Division of Pediatric Cardiology,\\
Children's Hospital of Philadelphia,\\
Philadelphia, PA 19104 \\
\texttt{wuw4@chop.edu} \\
\And
\textbf{Stephen Ching}\\
Department of Anesthesiology and\\
Critical Care Medicine,\\
Children's Hospital of Philadelphia,\\
Philadelphia, PA 19104 \\
\texttt{chings@chop.edu} \\
\And
\textbf{Patricia Sabin}\\
Department of Anesthesiology and\\
Critical Care Medicine,\\
Children's Hospital of Philadelphia,\\
Philadelphia, PA 19104\\
\texttt{sabinp@chop.edu} \\
\And
\textbf{Devin W. Laurence}\\
Department of Anesthesiology and\\
Critical Care Medicine,\\
Division of Pediatric Cardiology,\\
Children's Hospital of Philadelphia,\\
Philadelphia, PA 19104 \\
\texttt{laurenced@chop.edu} \\
\And
\textbf{Steve A. Maas}\\
Department of Biomedical Engineering, and\\
Scientific Computing and Imaging Institute, \\
University of Utah,\\
Salt Lake City, UT 84112\\
\texttt{steve.maas@utah.edu} \\  
\And
\textbf{Andras Lasso}\\
Laboratory for Percutaneous Surgery,\\
Queen's University,\\
Kingston, ON\\
\texttt{lasso@queensu.ca} \\  
\And
\textbf{Jeffrey A. Weiss}\\
Department of Biomedical Engineering, and\\
Scientific Computing and Imaging Institute, \\
University of Utah,\\
Salt Lake City, UT 84112\\
\texttt{jeff.weiss@utah.edu} \\  
\And
\textbf{Matthew A. Jolley}\\
Department of Anesthesiology and\\ 
Critical Care Medicine,\\
Division of Pediatric Cardiology,\\
Children's Hospital of Philadelphia,\\
Philadelphia, PA 19104\\
\texttt{jolleym@chop.edu} \\  
}

\begin{document}
\maketitle

\begin{abstract}
Advances in three-dimensional imaging provide the ability to construct and analyze finite element (FE) models to evaluate the biomechanical behavior and function of atrioventricular valves. However, while obtaining patient-specific valve geometry is now possible, non-invasive measurement of patient-specific leaflet material properties remains nearly impossible. Both valve geometry and tissue properties play a significant role in governing valve dynamics, leading to the central question of whether clinically relevant insights can be attained from FE analysis of atrioventricular valves without precise knowledge of tissue properties. As such we investigated 1) the influence of tissue extensibility and 2) the effects of constitutive model parameters and leaflet thickness on simulated valve function and mechanics. We compared metrics of valve function (\textit{e.g.,} leaflet coaptation and regurgitant orifice area) and mechanics (\textit{e.g.,} stress and strain) across one normal and three regurgitant mitral valve (MV) models with common mechanisms of regurgitation (annular dilation, leaflet prolapse, leaflet tethering) of both moderate and severe degree. We developed a novel fully-automated approach to accurately quantify regurgitant orifice areas of complex valve geometries. We found that the relative ordering of the mechanical and functional metrics was maintained across a group of valves using material properties up to 15\% softer than the representative adult mitral constitutive model. Our findings suggest that FE simulations can be used to qualitatively compare how differences and alterations in valve structure affect relative atrioventricular valve function even in populations where material properties are not precisely known.
\end{abstract}


\section{Introduction}

Recent advances in cardiac imaging, such as 3D echocardiography (3DE), offer a noninvasive means to evaluate the structure of dysfunctional heart valves \cite{Qureshi2019, muraru2019}. 3D images allow intuitive visualization of valve anatomy, identification of mechanisms of valve dysfunction, and quantification of the size of the valve leak (\textit{e.g.,} regurgitant orifice area (ROA)). In addition, 3D images enable the association of quantitative metrics of 3D valve structure to valvular function (degree of regurgitation) \cite{Utsunomiya2019, Oguz2019, Denisa2020, Karagodin2020, nam2022, Williams2022}, which in turn may inform valve repair \cite{Nguyen2019, Addetia2019, Jelena2022, Annachiara2022, Khabbaz2013}. However, association cannot infer causation, and it is not feasible to perform a randomized controlled trial of each proposed valve intervention due to the heterogeneity of clinical populations and the duration and cost to complete such studies \cite{Sacks2019}.  

3D image-derived computational approaches have the potential to meet the need for valve structural assessment and identification of the optimal repair for an individual patient. Over the last 20 years, extensive progress has been made in FE modeling of the mitral valve (MV), and more recently, the tricuspid valve \cite{Lee2014, Khalighi2017, Kamensky2018, Kong2018, Sacks2019, Singh-Gryzbon2019}. These investigations have used 3D image-derived models to investigate how variations in valve structure influence valve closure as well as leaflet stress and strain in adult atrioventricular valves \cite{Biffi2019, Sacks2019, Laurence2020, Wu2022}. In addition, a model of a dysfunctional valve can be realistically altered using computer-aided design (CAD) and FE analysis to precisely simulate the effect of adult atrioventricular valve repairs such as annuloplasty, leaflet resection, papillary muscle redirection, addition of prosthetic chords, and transcatheter edge-to-edge repair \cite{Choi2014, Choi2017, Kong2018_1, kong2020, Kamensky2018, Lee2019, Johnson2021}. These seminal works provide a roadmap for the controlled and reproducible virtual manipulation of leaflet structure, annular structure, and papillary muscle head position beginning from an initial valve model. FE simulations can then be applied to determine the effect of variations in both native valve anatomy and virtual interventions on leaflet coaptation, stress and strain, relative to a baseline valve model. 

\begin{figure}[h!]
\centering
\includegraphics[width=1\textwidth]{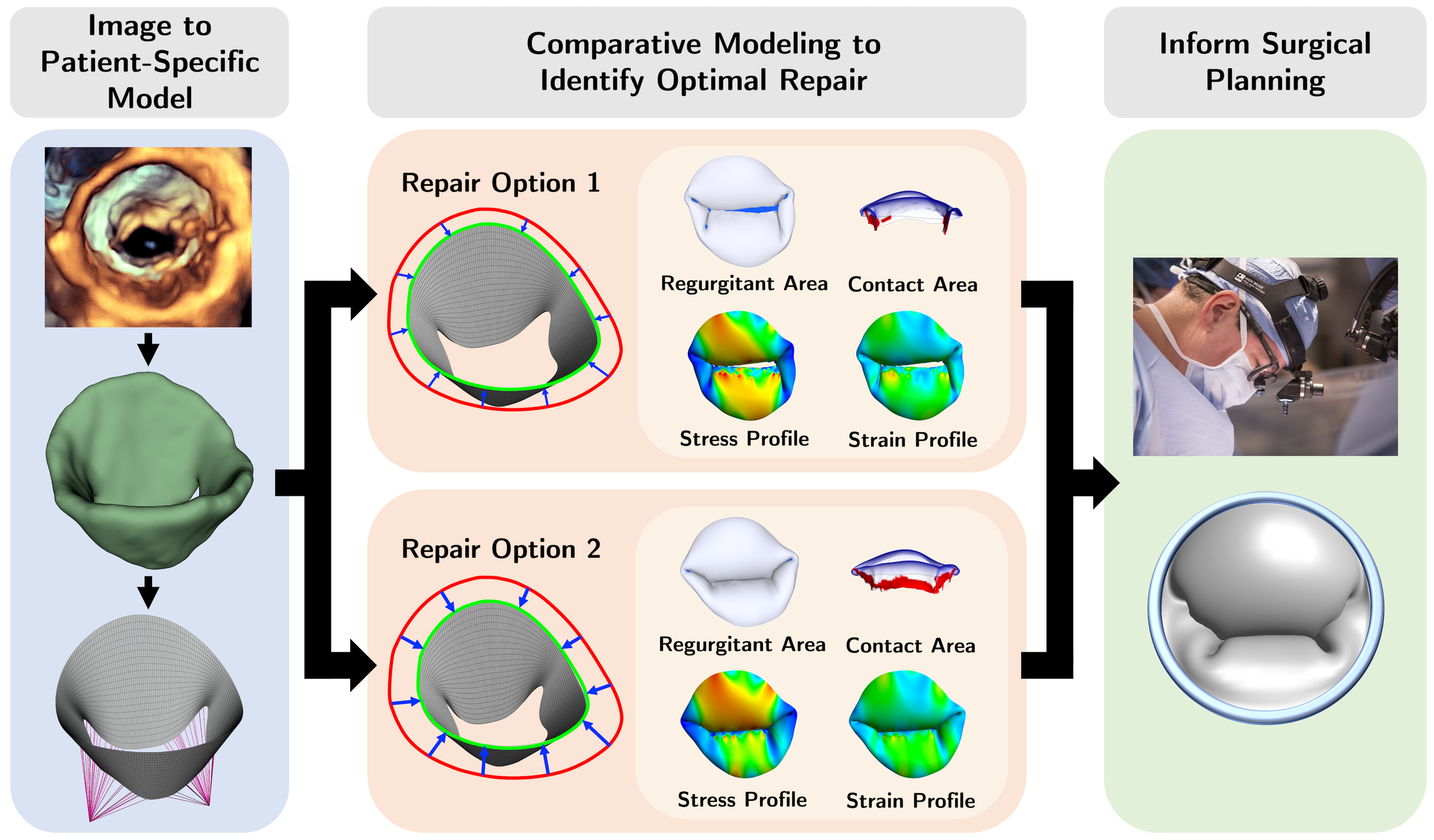}
\caption{\textbf{Conceptual future application workflow.} Potential future application workflow using image-derived simulation to explore and optimize the optimal repair for an individual patient. The workflow goes as follows 1) create a patient-specific 3D valve model from 3D image, 2) perform variations of virtual intervention to the valve model, and 3) identify the optimal repair by comparing the biomechanical and functional metrics for each option.}\label{workflow}
\end{figure}

While simulation of customized valve repairs is promising, patient-specific leaflet material properties cannot yet be noninvasively obtained, and the effect of variation in material properties on metrics of repair quality (leaflet coaptation, stress and strain) remains unknown. Notably, material properties may vary across individuals, and most modeling is currently performed using material parameters derived from animal or adult cadaveric tissue \cite{Sacks2019, Mathur2019, Wang2013, Pham2017}. Such variation is particularly relevant to the application of valve modeling in children, where valve leaflet material properties have not been described, and may differ across age and disease. Given both valve geometry and material properties are critical factors affecting leaflet mechanics and valve function \cite {Sacks2009, Hasan2014, Khalighi2017, Sacks2019, Laurence2020}, if the effect of variation in material properties is markedly greater than that of geometry, then evolving computationally derived optimizations of valve repair are less likely to be clinically informative. Specifically, this finding would compromise the otherwise promising development of image-derived computational modeling to inform atrioventricular valve repair until population or patient-specific material properties can be determined noninvasively. A schematic demonstrating the potential future application of image-derived finite element analysis and multi-physics simulation to inform surgical decision making in valve repair is shown in Fig. \ref{workflow}.

As such, we examined the effect of material properties on functional (leaflet coaptation area and ROA) and biomechanical (leaflet stress and strain) metrics across one normal (non-regurgitant) MV and six regurgitant MV model variants with different mechanisms of primary regurgitation characterized by abnormalities of the MV apparatus. We explored the influence of material properties on the relative magnitude of functional and biomechanical metrics across this range of valve geometry. We hypothesized that the relative ordering of the magnitude of functional and mechanical metrics of a group of valve models would be preserved across a range of leaflet material properties. This in turn would suggest that meaningful relative comparisons of valve function and mechanics could be obtained across variations of valve geometry (\textit{e.g.,} different interventions to repair a valve) even in the absence of precise knowledge of tissue material properties.

\section{Methods}
\begin{figure}[htbp]
\centering
\includegraphics[width=1\textwidth]{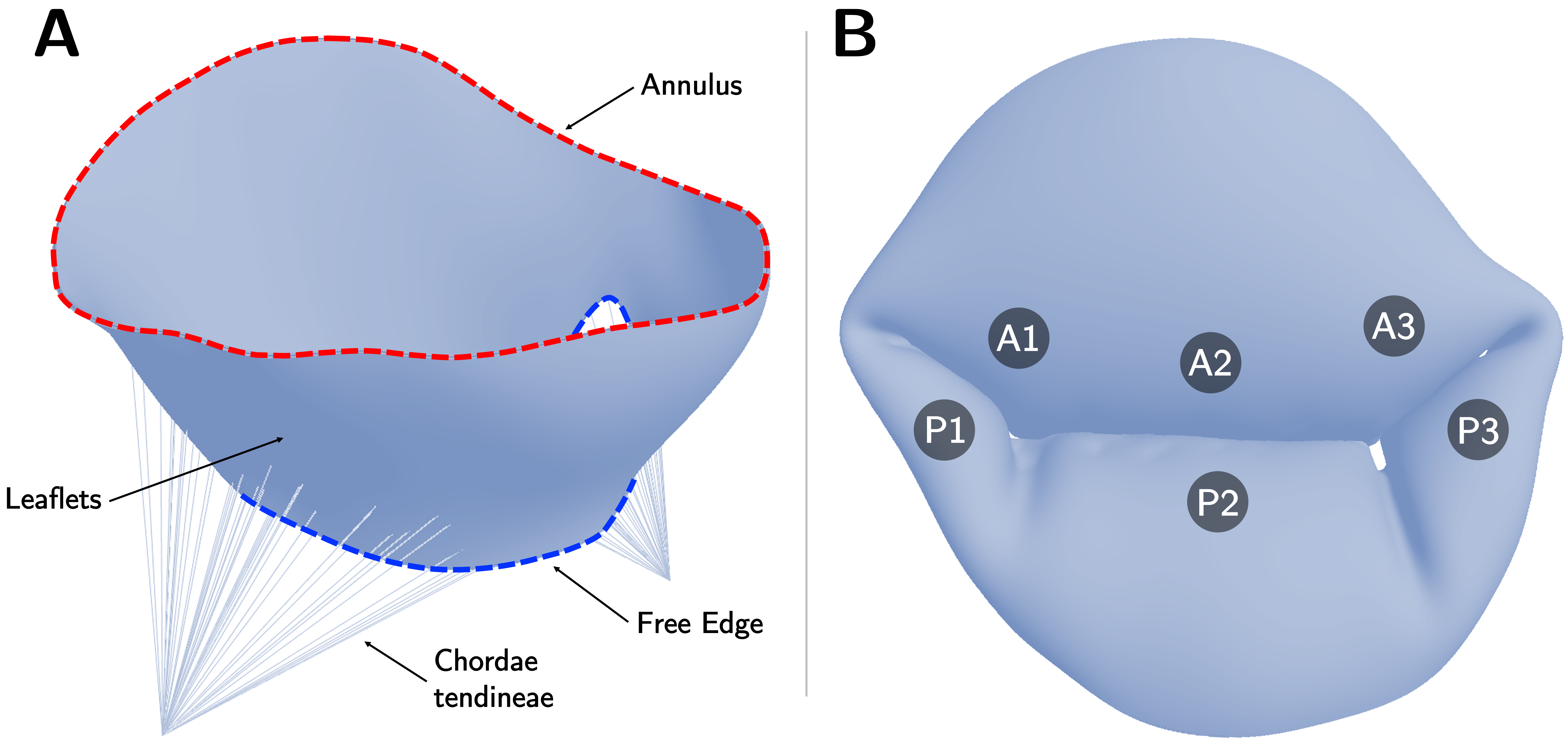}
\caption{\textbf{Mitral valve geometry and anatomy.} (\textbf{A}) Open mitral valve with annulus, free edge, leaflets, and chordae tendineae shown; (\textbf{B}) Closed mitral valve with regions of the anterior leaflet (A1, A2, A3) and posterior leaflet(P1, P2, P3) labeled.}\label{mitralNomenclature}
\end{figure}

\begin{figure}[h!]
\centering
\includegraphics[width=1\textwidth]{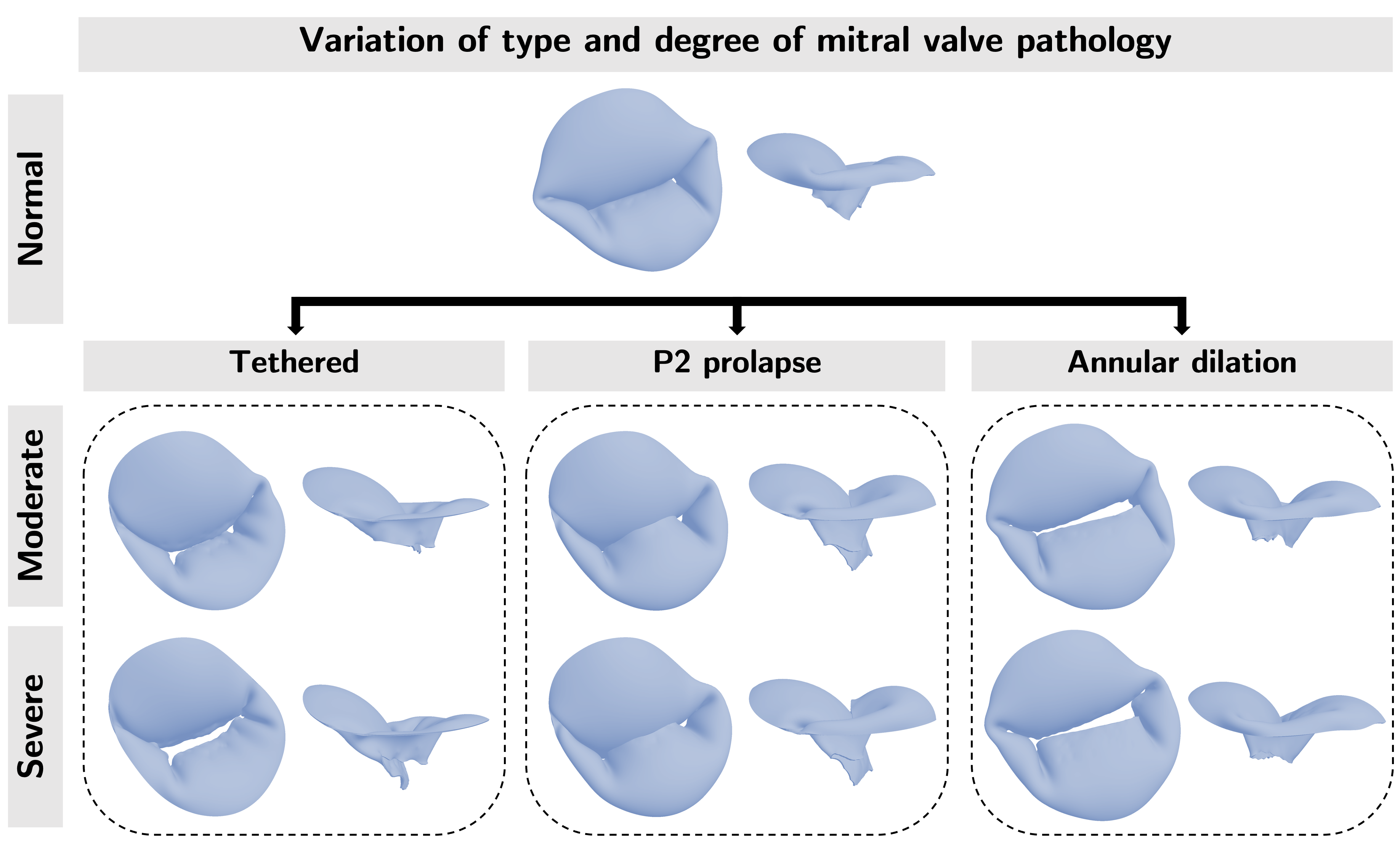}
\caption{\textbf{Overview.} We created models of one normal (non-regurgitant) and three regurgitant MV with different mechanisms of regurgitation ( posterior leaflet tethering, posterior leaflet prolapse in the P2 region, and symmetric annular dilation). For the three dysfunctional morphologies (Tethered, P2 prolapse, Annular dilation) we created morphologies with two degrees of regurgitation (moderate and severe). We used these MV morphologies to examine the effects of tissue extensibility and the individual material coefficient on the mechanical and functional metrics. Visualization of chordae omitted for clarity.}\label{overview}
\end{figure}

We aimed to evaluate the effects of varying soft tissue properties on mechanical and functional metrics of dysfunctional MVs using FE analysis. MV anatomy is shown in Fig. \ref{mitralNomenclature}. We first explored the influence of tissue extensibility on the mechanical and functional metrics of MVs with common mechanisms of regurgitation (annular dilation, posterior leaflet prolapse in the P2 region, and posterior leaflet tethering) across varied grades of regurgitation. Subsequently, we studied the effects of individual material parameters and leaflet thickness on the mechanical and functional metrics across severely regurgitant MVs. Fig. \ref{overview} provides an overview of the valve types considered in the present work. The FE model creation and analysis procedure are summarized in the following subsections. Interested readers may refer to Wu \textit{et al}. \cite{Wu2022} for more details regarding the computational modeling framework. 

\subsection{Finite element modeling procedure}

The computational analyses in the present work leveraged three open-source platforms: SlicerHeart \footnote{github.com/SlicerHeart} \cite{Nguyen2019, Lasso2022}, 3D Slicer \footnote{www.slicer.org} \cite{Fedorov2012}, and FEBio \footnote{www.febio.org} \cite{Maas2012, Maas2017, Ateshian2018}. Use of 3DE images for this work was approved by the Institutional Review Board at the Children's Hospital of Philadelphia.

To create a baseline FE mesh emulating realistic MV geometry, we used SlicerHeart to segment the valve leaflets of a normal adolescent male's MV from 3DE images. Then, we defined periodic control point splines at the annulus and free edge to extract the medial surface from the valve segmentation. Finally, we lofted a NURBS grid surface between the splines and fitted the grid surface to the valve medial surface. From this baseline geometry, the NURBS grid surface was edited to create a normal MV with full coaptation upon closure in FEBio. Three abnormal MV variants were created with further adjustments to the NURBS grid surface, papillary muscle tip placement, and chordae tension: (1) for posterior leaflet tethering, chordae length was increased by lowering papillary muscle tips and chordae tension was increased, (2) for posterior leaflet prolapse, chordae tension was decreased in the P2 region, and (3) for annular dilation, the diameter of the annulus curve was increased without lengthening the leaflets. For each of these three morphologies, models with two degrees of regurgitation severity (moderate, severe) were created.

The mechanical behaviors of leaflets are complex with properties contingent on collagen fiber concentration, characteristics, and orientation~\cite{Alavi2013}. Further, the collagen fiber distribution and organization variate across the four tissue layers within the leaflet structure~\cite{Goodwin2021}. In general, leaflet tissue exhibits an anisotropic behavior with the collagen fibers oriented with the circumferential direction, but in some cases, leaflets can be stiffer in the radial direction~\cite{Sadeghinia2022, Pham2014}. However, prior studies have demonstrated that leaflet anisotropy has negligible effects on the global leaflet deformation response~\cite{Wu2018}. As such, we adopted the incompressible, isotropic, hyperelastic Lee-Sacks constitutive model~\cite{Lee2014} to model the MV leaflet tissue for its simplicity and efficient formulation. The Lee-Sacks model is characterized by an unconstrained constitutive model in which the contributions of the extracellular matrix and collagen fiber network were approximated by combining neo-Hookean and exponential terms. Details of the constitutive model formulation can be found in~\cite{Lee2014, Kamensky2018}.

The FE meshes were discretized into 4-node linear quadrilateral (Quad4) shell elements \cite{Hou2018}. Pinned boundary conditions were applied to the annulus edge and the papillary muscle tips to restrict spatial translations in the valve models. A systolic pressure (100 mmHg) was prescribed to the ventricular surface of the leaflets to simulate valve closure. The leaflet contact was modeled using a potential-based formulation \cite{Kamensky2018}. The chords were modeled as previously described \cite{Wu2022}. Dynamic analyses were performed on the valve models using an implicit Newmark time integration scheme.

\subsection{Mechanical and functional metrics}

The mechanical and functional metrics considered include the average $1^\text{st}$ principal stress and strain, contact area(CA), and regurgitant orifice area (ROA). ROA (the orifice that blood can leak through from the ventricle to the atrium during ventricular contraction) and coaptation area are indices of immediate valve function (degree of regurgitation). The first principal stress and strain are measures that connect valve function (and pathology) to the tissue mechanical environment and are increasingly thought to be related to longitudinal valve function~\cite{El-Tallawi2021, Narang2021, marsan2021}. Recent studies in adult mitral valves demonstrate that greater leaflet stress and strain are associated with moderate or greater regurgitation, clinical valve repair failure, and pathologic changes in valve leaflets. 

The contact area for the Kamensky contact~\cite{Kamensky2018} was calculated by integrating the contact surface over the elements with non-zero traction applied to its integration points based on a force-separation law defined in Kamensky \textit{et al}.~\cite{Kamensky2018}. We developed a new fully-automated method to accurately quantify the ROA by coupling a shrink-wrapping method \cite{kobbelt1999, Overveld2004, weidert2020} with raycasting. This novel technique can capture multiple regurgitant orifices, and we have released it open-source in the Orifice Area module of SlicerHeart extension for 3D Slicer \cite{Lasso2022}. The procedures are shown in Fig. \ref{shrinkWrapMethod}. We included a tutorial video in the Supplement to demonstrate the pipeline for obtaining the biomechanical and functional metrics. In summary, we first imported the systolic frame of the valve FE model to 3D Slicer and inflated the surface mesh by the leaflet thickness. This step helps improve the accuracy of the quantification of the orifice opening. Second, we placed landmark points near the orifice opening to create a closed contour (Fig. \ref{shrinkWrapMethod}A). Third, we warped a flat disk template to the closed contour to generate a continuous surface \cite{Vigil2021} represented the red and blue patches combined in Fig. \ref{shrinkWrapMethod}B-D. The red patches indicated areas where the continuous surface is in contact with the leaflet surface, whereas the blue patches represented the potential orifice surface. Fourth, we iteratively smoothed (via Laplacian method), remeshed (via Approximated Centroidal Voronoi Diagrams algorithm \cite{Sebastien2004}), and displaced (in the direction of gradient vectors determined from the leaflet surface) the continuous surface towards the orifice opening, until the change in orifice surface area between iterations fell within a user-defined tolerance. The shrink-wrapping method may overestimate the potential orifice area in cases where the regurgitant opening is situated in a narrow and deep crease between the anterior and posterior leaflets. As such, we generated 400 rays in the surface normal direction of a 50-degree cone over the potential orifice surfaces to filter out invalid patches (Fig. \ref{shrinkWrapMethod}F). In other words, any patches with rays intersecting the leaflet surface were considered invalid orifice areas. Finally, we summed up all valid patches as the final, confirmed ROAs (Fig. \ref{shrinkWrapMethod}F). From our experience, we found 400 rays from each patch are sufficient for identifying small openings within reasonable computational time. Users may use fewer rays if desired. 
\begin{figure}[htbp]
\centering
\includegraphics[width=0.8\textwidth]{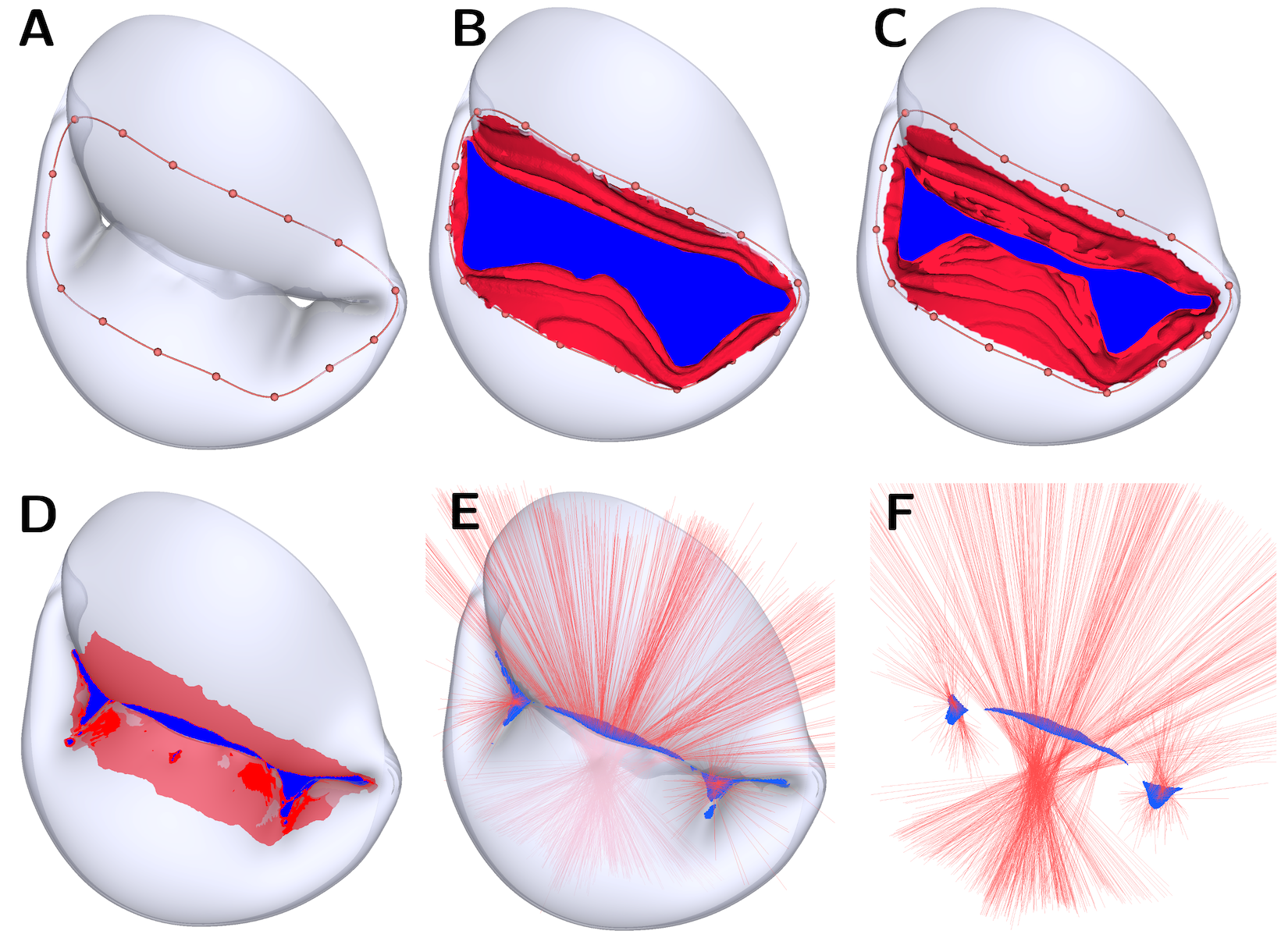}
\caption{\textbf{ROA computation procedure.} (\textbf{A}) Create a closed contour near the valve opening. (\textbf{B}) SlicerHeart automatically generates a continuous surface. The red region indicates the area where the continuous surface is in contact with the leaflet surface; blue indicates the potential orifice surface. (\textbf{C} to \textbf{D}) the potential orifice surface gradually descends toward the orifice opening in the valve model by a user-defined number of shrink-wrapping iterations. (\textbf{E}) Generate 400 rays (red lines) over the potential orifice surface to identify the streamlines that pass through the valve without intersecting the leaflets. (\textbf{F}) Compute the effective ROA by summing the areas around the rays that do not intersect the leaflet surface.}\label{shrinkWrapMethod}
\end{figure}

\subsection{Numerical derivation of material variants}

\begin{figure}[htbp]
\centering
\includegraphics[width=1\textwidth]{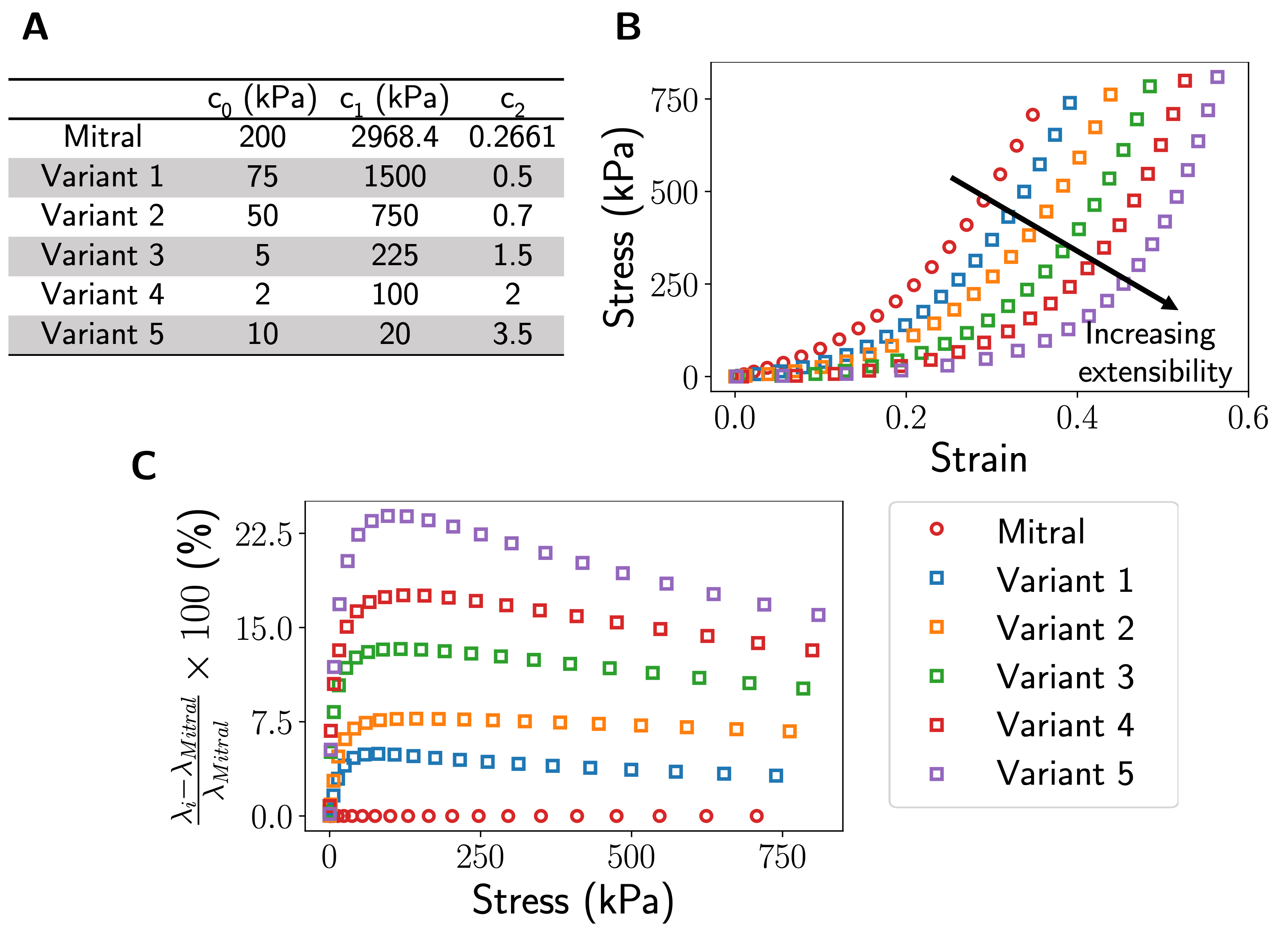}
\caption{\textbf{Material variants.} We derived five material variants, each representing a gradual increase of tissue extensibility relative to the baseline mitral model, by performing a uniaxial test using FE analysis. The material constants are shown in (\textbf{A}), their corresponding stress-strain curves are shown in (\textbf{B}), and the percentage difference of the stretch ratio, $\lambda$, between each material variant and the mitral model (\textbf{C}).}\label{material}
\end{figure}

While there are representative material models to capture leaflet tissue properties in adults, characterizing tissue properties in children remains a challenging task due to the inaccessibility of human tissue. This limitation hinders meaningful studies of valve functions and closure. As many studies suggested that heart valve tissue is softer in children \cite{Stephens2010, Kasyanov2013, Geemen2016, Wu2020}, we performed uniaxial FE analysis on a single-element FE model and numerically derived five sets of material coefficients that yielded various degrees of increased extensibility from the adult mitral leaflet properties (Fig. \ref{material}). These models were used to represent the range of leaflet properties in children until more precise tissue material properties are available. 

\subsection{Uncertainty analysis}

In addition to evaluating the influence of tissue extensibility on the mechanical and functional metrics of a diverse set of regurgitant MVs, we also examined the sensitivity of material coefficients ($c_0$, $c_1$, and $c_2$ in the Lee-Sack constitutive model) and leaflet thickness on the mechanical and functional outcomes of the normal and severely regurgitant MVs using 1) a traditional approach and 2) a statistical approach. In the traditional approach, for each mitral pathology, we uniformly sampled five data points in the range of $\pm 50\%$ of the reference value per model parameter (\textit{i.e.,} a total of 20 FE models were considered per valve model). Then, we performed FE analyses by varying one material coefficient at a time, while keeping the reference values for the remaining coefficients constant to identify the sensitivity of the mechanical and functional metrics with respect to individual material model coefficients. While traditional uncertainty analysis provides a systematic approach for evaluating model sensitivity with respect to uncertainty in individual input parameters, the statistical approach facilitates a more comprehensive assessment of the model outcomes from uncertainty across multiple parameters. The latter approach may be a more suitable choice for our application since the input parameters at play (material coefficients and tissue thickness) are dependent on each other nonlinearly. For example, a change in $c_0$ would also necessitate a change in the other parameters to produce the same constitutive behavior. In the statistical approach, for each dysfunctional valve, we randomly generated 80 sets of input parameters in a 4-dimensional space, with each dimension spanning the range of $\pm 50\%$ of the reference value for each input parameter, using a polynomical chaos expansion function with a weighted approximate Fekete points (WAFP) method. The reference value for the input parameters and their ranges are shown in Table \ref{uncertainty_parameters}. This allowed us to assess the total-order sensitivity indices, a metric that measures the combined effect of each input parameter in a multi-parametric space from parameter interaction (the total contribution of multiple parameter variations)~\cite{Burk2020, Wu2022}. This statistical uncertainty analysis was achieved via a newly developed Python subroutine called FEBioUncertainSCI that interfaces UncertainSCI \cite{Burk2020} with the FEBio solver. The source code of the subroutine can be found at \url{github.com/febiosoftware/FEBioUncertainSCI}. 

\begin{table}[htbp] 
\centering
\caption{\textbf{Uncertainty analysis sample points.} We sampled a total of five points uniformly between the minimum and maximum values for each parameter to identify the sensitivity of the mechanical and functional metrics with perturbation in the material constants and leaflet thickness.}
\begin{tabular}{c}
\includegraphics[width=0.7\textwidth]{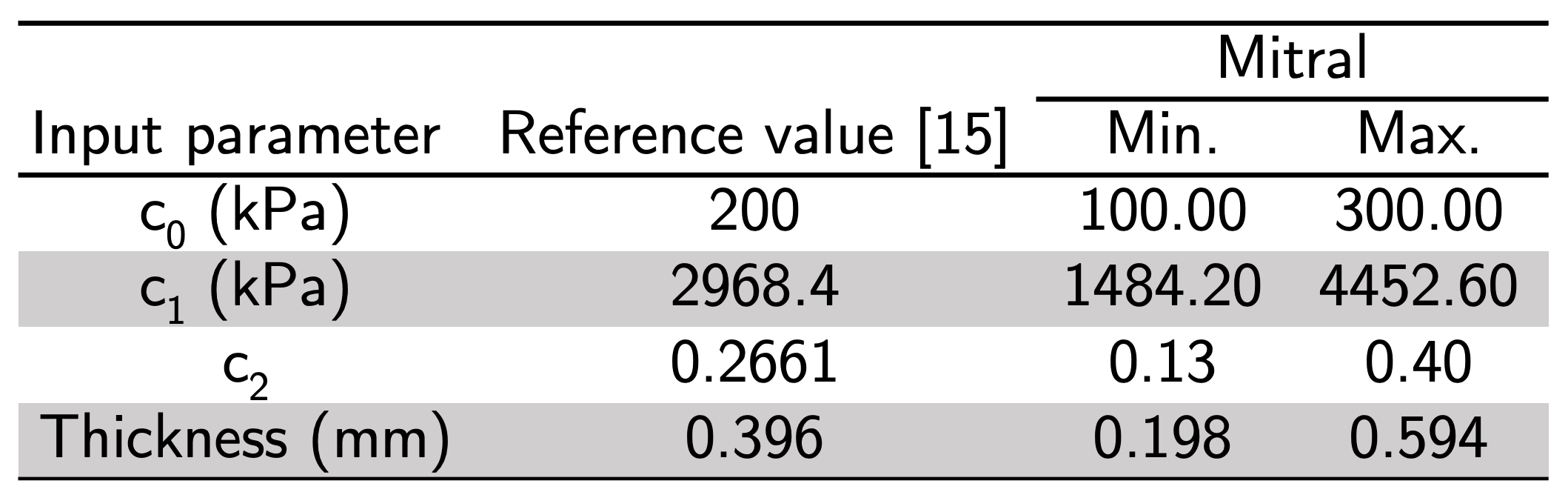} 
\end{tabular}\label{uncertainty_parameters} 
\end{table}

\section{Results}

In Section \ref{parametric_analysis}, we altered material extensibility to examine the effects of tissue stretches on the mechanical and functional metrics for pathologic valves. In Section \ref{uncertainty_analysis}, we performed an uncertainty analysis for one normal and three severely regurgitant valves to discern the effects of individual material model parameters and the corresponding mechanical and functional metrics at valve closure.

\subsection{Mechanical and functional metrics comparative studies} \label{parametric_analysis}

\subsubsection{Similar regurgitant MV geometries}

We applied five sets of material coefficients to regurgitant valves of moderate and severe regurgitant grades to assess the influence of tissue extensibility on the mechanical and functional metrics in MVs of similar geometries. We report our findings for the tethered MV in this section to save space; readers may refer to Appendix \ref{appendix:a} for the remaining findings.

\begin{figure}[htbp]
\centering
\includegraphics[width=1\textwidth]{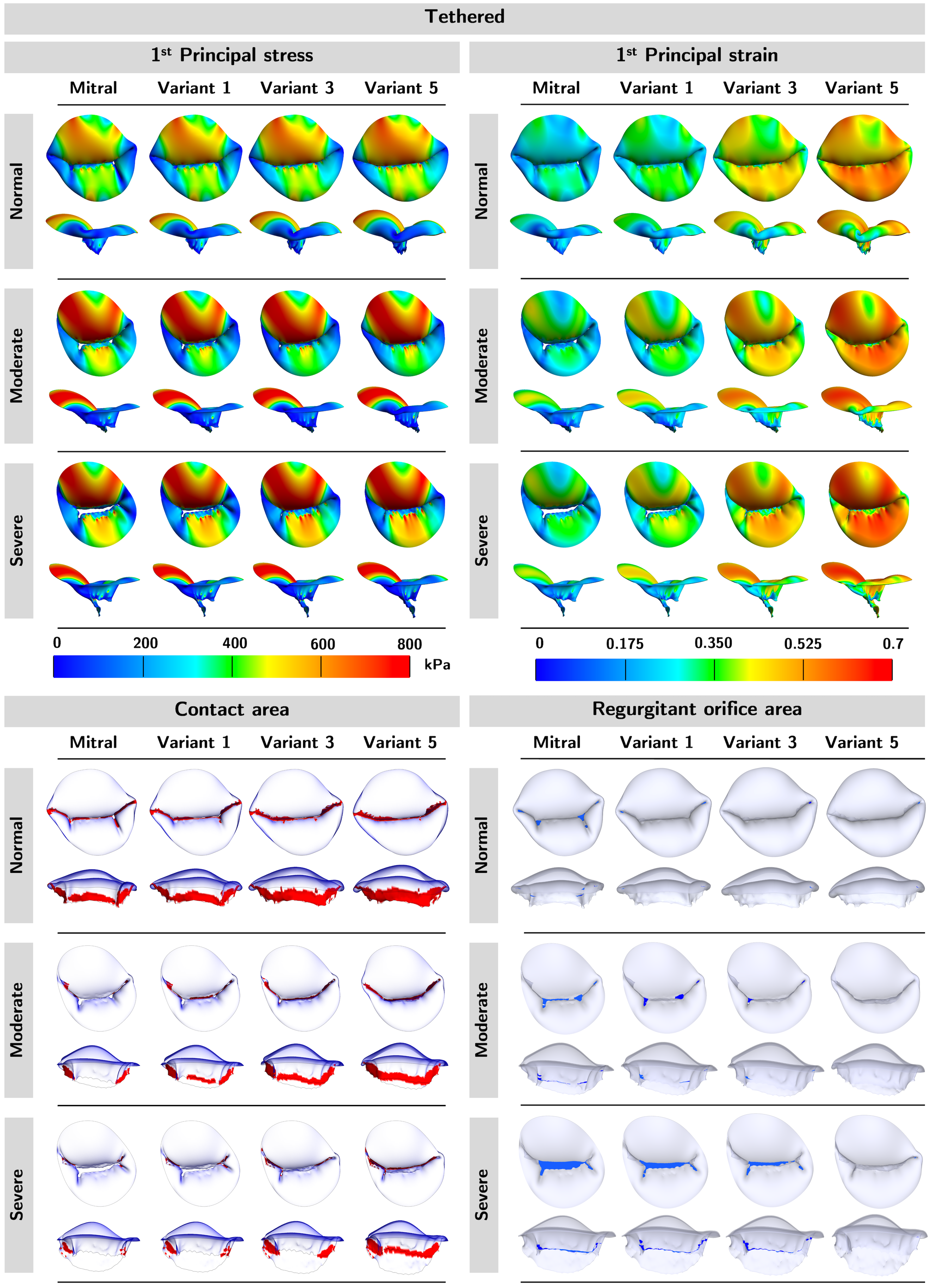}
\caption{\textbf{Mechanical and functional metric profiles for tethered models.} We observed an increase in $1^\text{st}$ principal stress, $1^\text{st}$ principal strain, and contact area but a decrease in ROA in the softer material variants. Additionally, valves with severe regurgitation had higher stress, strain, and regurgitant area, but less contact area.} \label{tetheredMetrics}
\end{figure}

The top and side profiles of mechanical and functional metrics for the tethered mitral are shown in Fig. \ref{tetheredMetrics}. We observed that the softer material coefficients led to higher 1$^\text{st}$ principal stress and strain, more contact area, and less ROA. In addition, as the regurgitation severity increased (\textit{i.e.,} ROA), so did the stress and strain. Again, the color red in the contact area panels represents the contact area, and the color blue in the ROA panels represents the ROA.

\begin{figure}[htbp]
\centering
\includegraphics[width=1\textwidth]{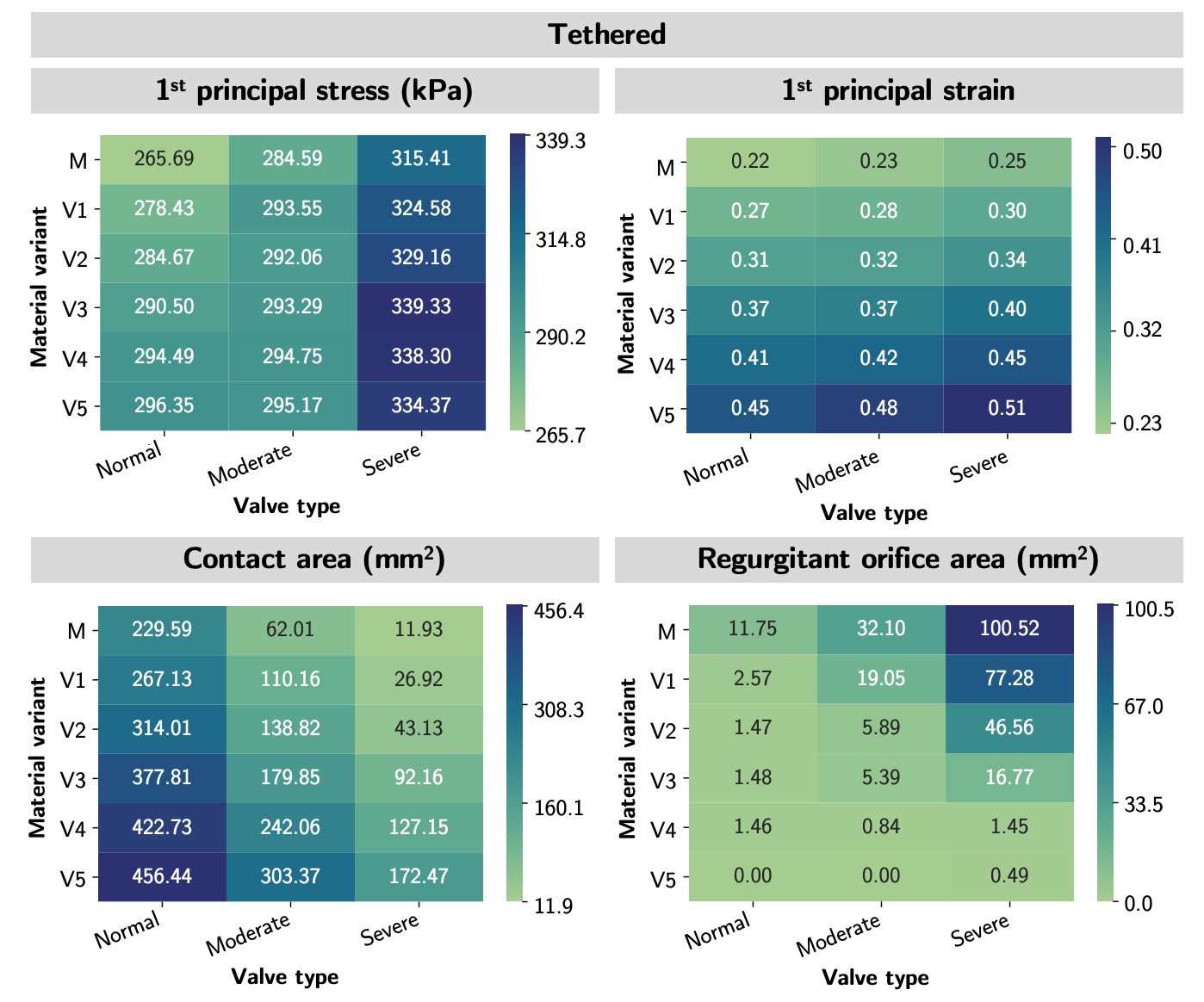}
\caption{\textbf{Mechanical and functional metric patterns for tethered models.} The heatmaps demonstrated the trend of the mechanical and functional metrics by a) regurgitation severity (across the columns) and 2) by material variant (across the rows). The relative ordering of the mechanical and functional metrics was mostly preserved in both scenarios. Material variants: M-Mitral; V1-Variant 1; V2-Variant 2; V3-Variant 3; V4-Variant 4; V5-Variant 5.} \label{tetheredMetricsRanking}
\end{figure}

In Fig. \ref{tetheredMetricsRanking}, we provided a heatmap for each mechanical and functional metric to illustrate the pattern of the metric by regurgitation severity (across the columns) and by material extensibility (across the rows). In both cases, the relative orderings across the mechanical and functional metrics were maintained, except for the first principal stress where negligible ordering inconsistency across regurgitant severity was observed. Altogether, this suggested that the relative ordering of the mechanical and functional metrics was maintained across diseased valves of similar geometries, for the range of material properties considered in the present work. Therefore, imprecise approximation of \textit{in vivo} leaflet tissue extensibility does not affect the validity of comparative mechanical and functional outcomes across MVs of similar geometries.

\subsubsection{Disparate regurgitant MV geometries}
\label{studies::severeRegurgitantMitral}
To understand whether valve geometries or material properties dominate the mechanical and functional metrics, we applied the five sets of material parameters to one normal and three severely regurgitated MVs (tethered, P2 prolapse, and annular dilation) representing a diverse range of MV geometries.

\begin{figure}[htbp]
\centering
\includegraphics[width=1\textwidth]{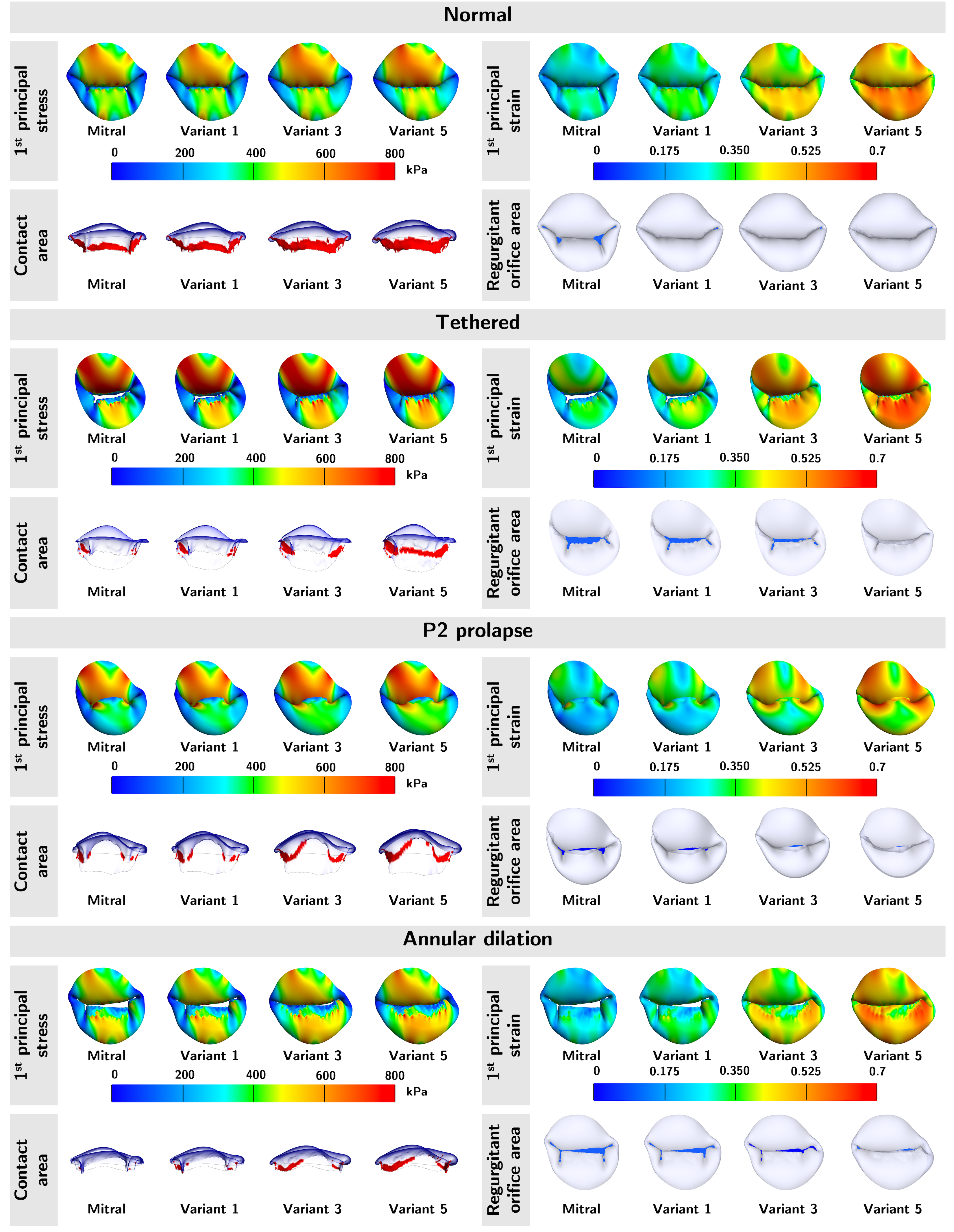}
\caption{\textbf{Mechanical and functional metric profiles for normal and severely regurgitant MVs.} The average $1^\text{st}$ principal stress, average $1^\text{st}$ principal strain, contact area, and ROA of the normal and three diseased MVs were shown. Variants 1, 3, and 5 represent tissue material variants shown in Fig. \ref{material}, with variant 1 being the stiffest and 5 being the softest among the numerically derived material variants. Across the four valves, the $1^\text{st}$ principal stress, $1^\text{st}$ principal strain, and contact area increased with softer tissue; the opposite for the ROA. Regarding stress/strain patterns, the highest stress and strain were found in the A1 and A3 regions of the P2 prolapse and tethered valves, and A1 region of the annular dilation valves.} \label{valvePathology}
\end{figure}

 The top and side views of average $1^\text{st}$ principal stress, average $1^\text{st}$ principal strain, contact area, and ROA for each of the abnormal valves are shown in Fig. \ref{valvePathology}. Note, the color red in the contact area panels represent the contact area, and the color blue in the ROA panels represents the ROA. Across the four MVs, we observed higher stress, strain, and contact area, but lower ROA in material models with higher extensibility. While the stress and strain patterns were consistent across different material models within each type of regurgitant valve, the stress and strain profiles were significantly different across the regurgitant valves. The stresses and strains were highest in the A1 and A3 regions of the P2 prolapse and tethered valves, and A1 region of the annular dilation valves. This inconsistency suggested that valve geometries (\textit{i.e.,} the annular geometry, leaflet geometry, and subvalvular apparatus) had a considerable effect on the stress and strain profiles of the valves. 

\begin{figure}[htbp]
\centering
\includegraphics[width=1\textwidth]{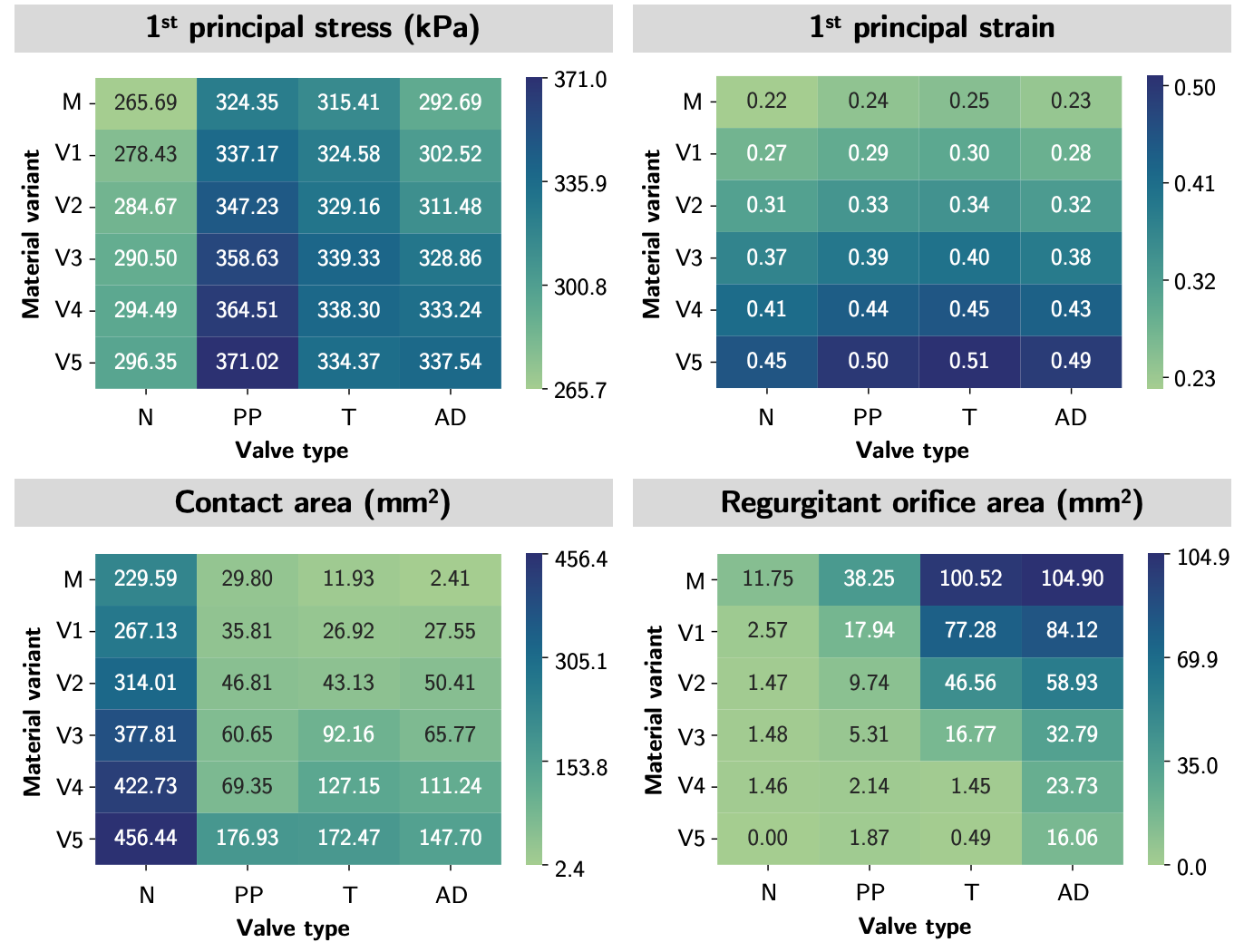}
\caption{\textbf{Mechanical and functional metric patterns for tethered models.} The heatmaps demonstrated the trend of the mechanical and functional metrics by a) regurgitation MVs (across the rows) and 2)by material variant (across the columns). The relative ordering of the mechanical and functional metrics was mostly preserved in both scenarios. Valve types: N-normal; PP-P2 prolapse; T-tethered; AD-annular dilation. Material variants: M-Mitral; V1-Variant 1; V2-Variant 2; V3-Variant 3; V4-Variant 4; V5-Variant 5.} \label{regurgitantMetricsRanking}
\end{figure}

In Fig. \ref{regurgitantMetricsRanking}, we provided a quantitative heatmap for each mechanical and functional metric to illustrate the pattern of the metric across mitral pathology (across the columns) and across material extensibility (across the rows). We observed excellent consistency in the patterns across valve pathology with tissue extensibility from mitral to Variant 3 in all metrics. In other words, the relative ordering of the metrics of valve mechanics and ROA were maintained for material models up to approximately 15\% more stretchable than the mitral material properties. This finding indicated that the comparative mechanical and functional outcomes were still consistent in MVs of significantly dissimilar geometries, so long the true \textit{in vivo} tissue softness was within 15\% of the representative MVs material parameters. Further, the relative ordering of the mechanical and functional metrics was mostly preserved across material variants. This finding was consistent with our intuition that softer tissue yields higher stresses, strains, and contact area ratio but a lower ROA. 

\subsection{Uncertainty analyses for dysfunctional MVs} \label{uncertainty_analysis}

\begin{figure}[htbp]
\centering
\includegraphics[width=1\textwidth]{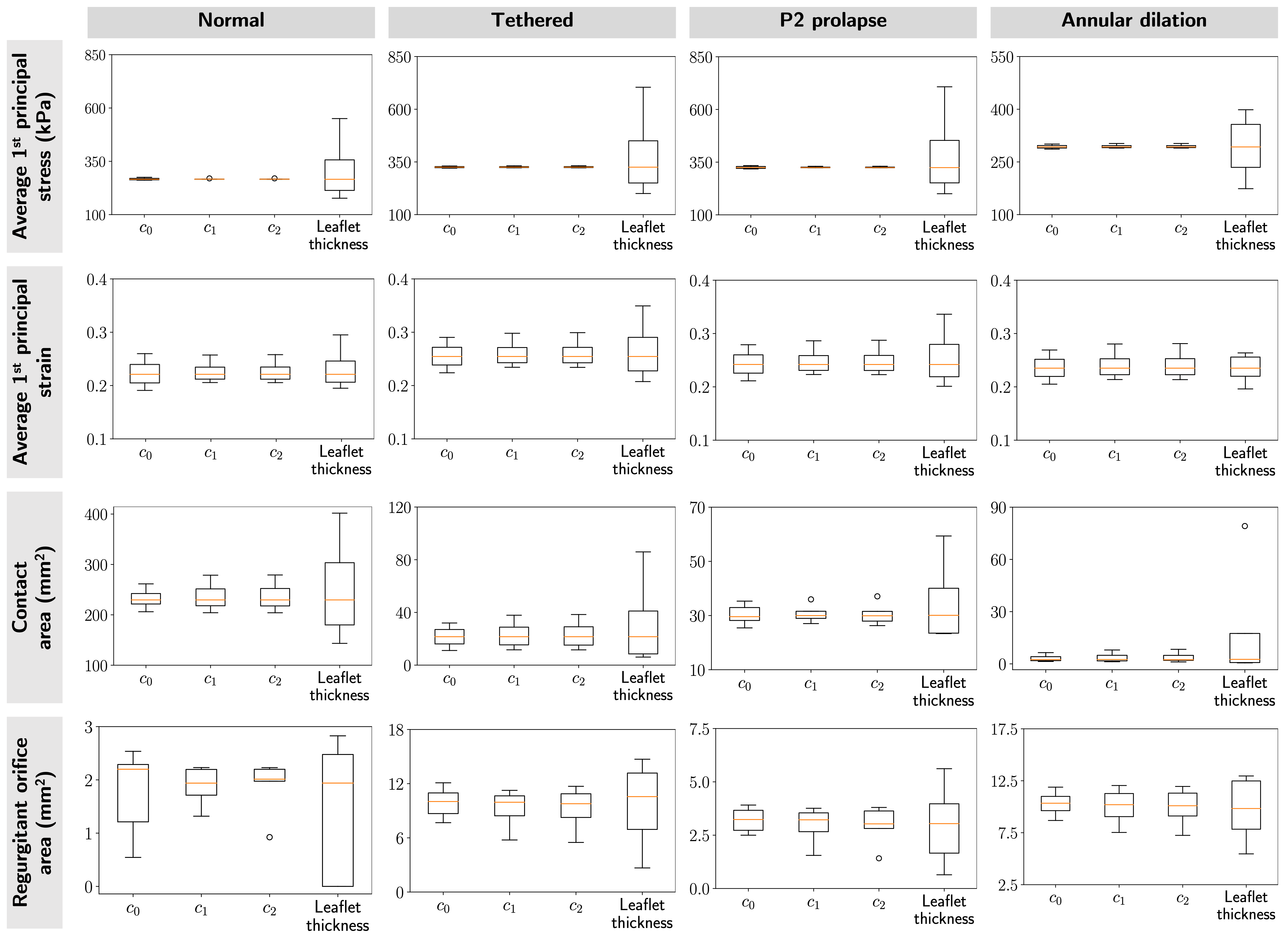}
\caption{\textbf{Traditional sensitivity results.} We evaluated the sensitivity of the mechanical and functional metrics to perturbation in model parameters for four valves: normal, tethered, P2 prolapse, and annular dilation. The box plots represented the spread of skewness in the mechanical and functional metrics for each modeling parameter. These results indicated the mechanical and functional metrics were most sensitive to leaflet thickness, consistent across the four valves. Additionally, all three material constants had a negligible effect on leaflet stresses.}\label{sensitivity}
\end{figure}

\begin{figure}[htbp]
\centering
\includegraphics[width=1\textwidth]{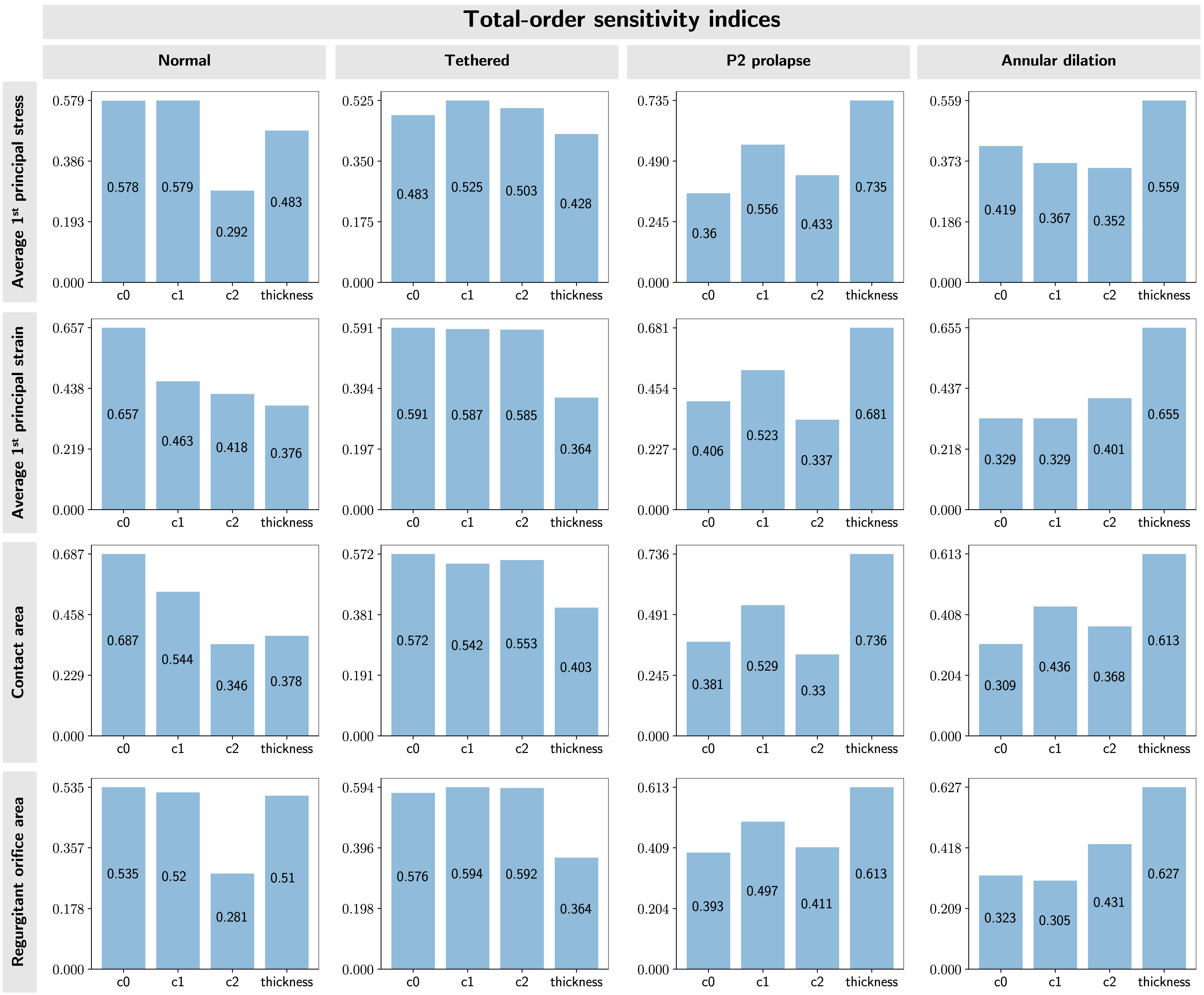}
\caption{\textbf{Statistical sensitivity results.} We evaluated total-sensitivity indices of the model input parameters to account for the effects of higher-order interactions between input parameters. These results indicated that all input parameters had an equal influence on the mechanical and functional metrics.}\label{sensitivity:uncertianSCIs}
\end{figure}

Fig. \ref{sensitivity} and \ref{sensitivity:uncertianSCIs} shows the effects of material coefficients ($c_0$, $c_1$, $c_2$) and leaflet thickness on the average $1^{\rm{st}}$ principal stress and strain at valve closure. In the traditional approach, where only one parameter was varied at a time, we observed that the material constants $c_0$, $c_1$, and $c_2$ had similar influences on the mechanical and functional metrics across the four valves. Further, the metrics were comparatively more sensitive to changes in leaflet thickness. The line plots that demonstrated changes in mechanical and functional metrics concerning model parameters can be found in Appendix \ref{appendix:b}. 

In our statistical approach (UncertainSCI), where multiple input parameters were varied simultaneously, the total sensitivity indices were similar across the input parameters across valve types. This result suggested that the parameter interactions had a significant contribution to the total sensitivity of the FE model, and provided a more thorough representation of the simulation results given the dependent nature of the material constants. This additional insight provided by the statistical approach highlights the importance of considering parameter interactions in sensitivity analysis.

\begin{figure}[htbp]
\centering
\includegraphics[width=1\textwidth]{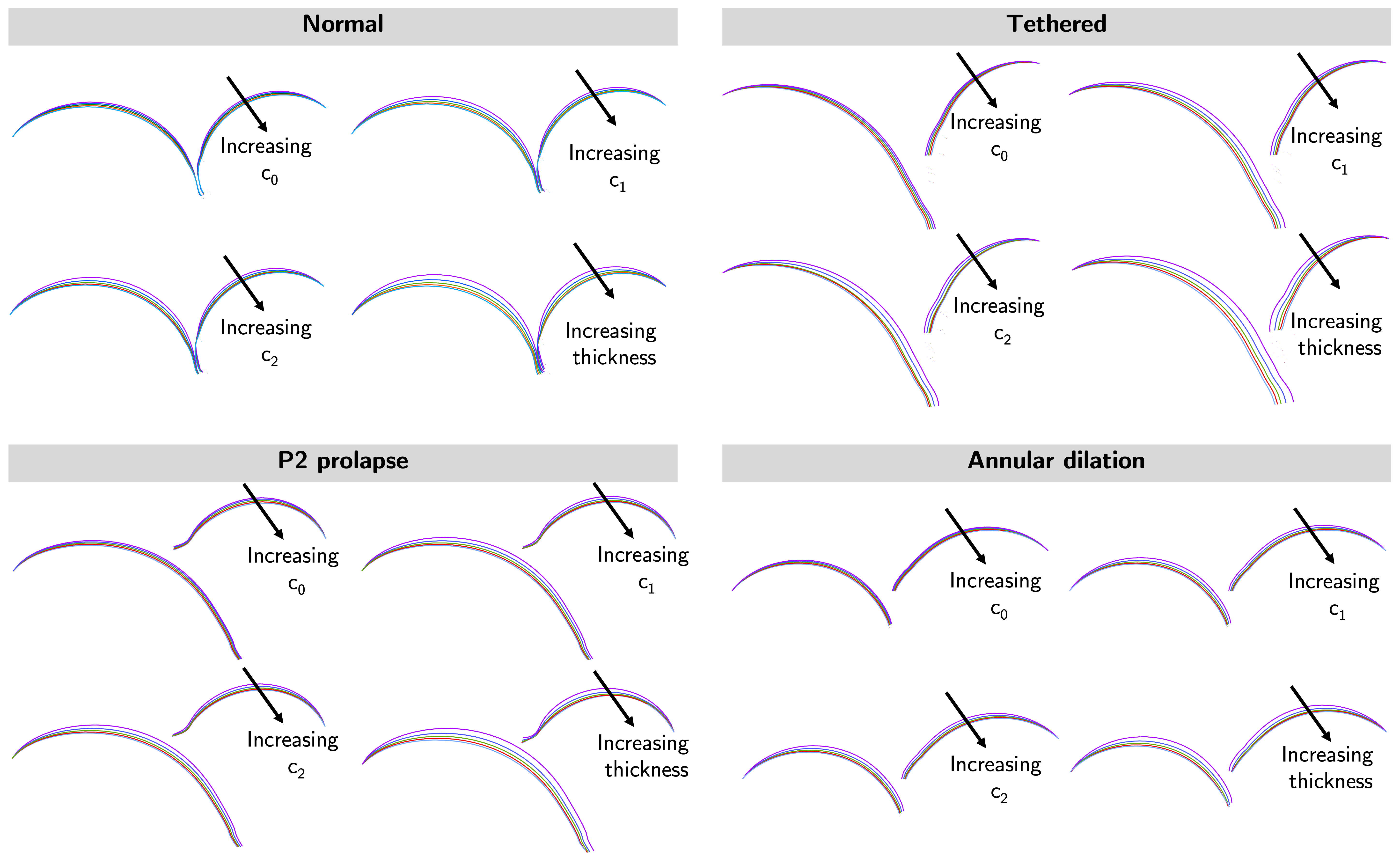}
\caption{\textbf{Cross-sectional views of mitral models in systole.} Cross-sectional views of the FE models were shown. The leaflet deformation was relatively insensitive to material coefficient $c_0$, moderately sensitive to material coefficients $c_1$ and $c_2$, and most sensitive to leaflet thickness.}\label{sliced}
\end{figure}

Fig. \ref{sliced} presents the sliced view of the dysfunctional valves to assess the effects of model parameters on leaflet deformation at systole. Our results indicated leaflet deformation was relatively insensitive to tissue coefficient $c_0$. However, noticeable differences in leaflet deformation were observed in situations with varying $c_1$, $c_2$, and leaflet thickness; leaflet deformation reduced with higher $c_1$, $c_2$, and thicker leaflets.  

\section{Discussion}
\subsection{Overall findings}
\label{Discussion:Findings}
In the present work, we sought to understand how uncertainty and variation in heart valve leaflet mechanical properties influence FE-derived predictions of valve function and mechanics. We found that the relative ordering of the magnitude of functional and mechanical metrics across a range of MV morphologies was consistent when a tissue extensibility within 15\% of the baseline model was employed. This suggests that geometry is the dominant determinant of metrics of valve function and mechanics within this range of variation in material properties. 

We explored how variations in tissue extensibility influenced FE metrics and comparisons across functional and dysfunctional MV morphology. As expected we noted that each individual MV geometry had similar stress and strain contours, with differing magnitudes, as constitutive model parameters were varied. Despite the clear influence of the gradual increase of tissue extensibility on the FE predictions, the relative ordering of the functional metrics was maintained across a range of leaflet material properties within 15\% of the baseline mitral model stretch ratio. Further, the relative ordering of functional metrics for dysfunctional valves with increasing regurgitation severity was preserved across the leaflet material properties defined in this work. These results suggest that meaningful qualitative insights into the immediate (\textit{i.e.,} functional metrics) and long-term (\textit{i.e.,} mechanical metrics) valve function can be obtained in the absence of precise knowledge of leaflet tissue properties. This finding  further supports the further development and validation of image-derived patient-specific computational modeling to inform the optimization of valve repair, as discussed further below.

We also explored the effects of uncertainty in tissue thickness and constitutive model parameters on the FE-derived valve metrics using traditional and statistical uncertainty analysis. In the traditional analysis, our results showed that the mechanical metrics (\textit{i.e.,} tissue stress and strain) and the functional metrics (\textit{i.e.,} valve contact area and ROA) were most sensitive to the tissue thickness. We noted that all three constitutive model parameters had a similar influence on the simulation results. These trends were consistent for all four MV geometries, but the magnitude of variation differed for each metric with tissue thickness having the most profound impact on the predicted leaflet stress. However, gleaned from statistical analysis results, it was inconclusive on which input parameter had the most dominant influence on the mechanical and functional metrics. We found that the leaflet thickness and the constitutive model parameters had similar total-order sensitivity indices, indicating that they had an equivalent influence on the relative effects of the mechanical and functional metrics in the FE analysis results. We noted that the statistical sensitivity analysis provided a more comprehensive representation of the model responses compared to traditional sensitivity analysis, as it took into account the combined effects from higher-order input parameter interactions. Although the statistical analysis results suggested that uncertainty in material constants had an influence on the resulting magnitude of each metric, this additional finding would not affect the overall conclusion on the relative ordering of the valve function and mechanics across normal and dysfunctional valves, as consistent constitutive parameters and material thickness were used when comparing metrics across valve types.

\subsection{Comparisons with existing literature}
Previous studies have used intricate animal models and experimental systems to determine heart valve leaflet mechanical properties for developing high-fidelity simulation tools. We noted that our predicted MV leaflet average stresses for the normal model ($265.69$~kPa) were within a similar range as the previous studies of Lee \textit{et al}.~\cite{Lee2014} ($240$-$600$~kPa) and Rim \textit{et al}.~\cite{Rim2013} ($150-300$~kPa). Furthermore, our predicted average principal strains ($0.220$) were within the range of previous experimental findings of Lee \textit{et al}.~\cite{Lee2015} and Amini \textit{et al}.~\cite{Amini2012} for porcine valves ($10$-$35\%$), but were larger than the strains determined by Rausch \textit{et al}.~\cite{Rausch2011} for sheep valves ($4$-$7\%$). Slight differences may be attributed to differing transvalvular pressures, constitutive model forms, assumed boundary conditions (\textit{e.g.,} dynamic annulus), and the lack of pre-strains.

There has been an emerging interest in using cardiac imaging to determine the functional mechanical properties of heart valves and better understand valve disease. Our predicted strains were larger than a recent clinical study that used cardiac imaging to determine functional MV leaflet strains of $7$-$9\%$~\cite{El-Tallawi2021}. However, our numerical results echoed their observation that diseased valves have larger strains than healthy valves. This suggests that underlying valve morphology is the most important determinant of valve deformation, and further illustrates that FE simulations may be used as an exploratory tool without precise \textit{a priori} knowledge of constitutive model parameters. Ultimately, this expands on the versatility of image-derived investigations, such as the recent work by Narang \textit{et al}.~\cite{Narang2021} that found differences in leaflet shear strains for adult MVs with recurrent ischemic mitral regurgitation.

\subsection{Clinical implications}
In this work, we sought to explore extending the application of FE methods to clinical applications (\textit{e.g.,} comparison of different repair strategies) where exact or patient-specific tissue material properties are unknown. Our results show that FE modeling predictions can be used to understand relative disease severity and the effect of alterations to the leaflet mechanical environment so long as material property parameters are maintained across the comparison group and fall within the range of $\sim15\%$ of the baseline model. For example, comparing different valve repairs for atrioventricular valve disease could significantly contribute to the optimization of valve repairs in small and heterogeneous populations less likely to benefit from iterative clinical refinement, such as children with congenital heart disease. Future inverse methods leveraging patient-specific images may inform these developments, but the present work has illustrated that FE analysis can be used to derive meaningful comparisons in parallel with the maturation of our understanding of material properties of atrioventricular valves, and in particular pediatric atrioventricular valves.

An additional contribution of this work is a new method we have implemented in SlicerHeart that couples a shrink-wrapping method \cite{kobbelt1999, Overveld2004, weidert2020} with raycasting to automatically and reproducibly determine the ROA of heart valves. Previous simulation studies relied on simplifying the closed valve geometry to two-dimensional slices~\cite{Laurence2020} or sampling different perspectives to find the largest orifice area~\cite{Johnson2021}. Although both methods may provide reasonable approximations of the ROA, they are orientation dependent, may fail to accurately capture the complex closed valve geometry (\textit{e.g.,} multiple jets), and consequently often misrepresent the true ROA. Our methodology overcomes these limitations by employing an algorithm that is invariant to the view angle and can consider multiple regurgitant jets across the valve. 

\subsection{Study limitations and future directions}
\label{Discussion:Limitations}
This study includes several convenient simplifications on the FE model representations of the MV sub-valvular apparatus and leaflet mechanical properties that do not detract from the quality of the findings. In terms of FE model simplifications, first, we assumed pinned boundary conditions for the annulus and papillary muscle tips rather than dynamic, from ventricular contraction and relaxation, as they would be \textit{in vivo}. Considering dynamic annular and papillary motion in the FE simulation may reduce the predicted leaflet stresses~\cite{Rim2013}, but this will not affect our comparisons as all FE simulations share this assumption. Second, we modeled each of the anterolateral and posteromedial papillary muscles with a single tip and assumed uniform chordae distribution over an anatomic area of chordal insertion extending inward from the leaflet edge. However, chordae and papillary muscle anatomy may differ in patients with functional and degenerative mitral regurgitation~\cite{Obase2014, Rajiah2019}. Considering patient-specific sub-valvular components may provide a more accurate assessment of the functional and biomechanical metrics when such imaging becomes clinically feasible. 

We used an isotropic form of the Lee-Sacks constitutive model with adult MV parameters that cannot capture the typical anisotropic mechanical behaviors of heart valve leaflets. Fortunately, Wu \textit{et al}.~\cite{Wu2018} demonstrated that this anisotropy plays a minimal role in predicting large-scale tissue deformations and subsequent geometry-based functional metrics (\textit{e.g.,} contact area). Further, it is known that the heart valve leaflets have regionally heterogeneous properties (\textit{e.g.,} thickness~\cite{Lee2015} and stiffness~\cite{Fitzpatrick2022}). Consideration of leaflet anisotropy and heterogeneous thickness may confound comparisons between healthy and diseased valve function. Notably, the valve leaflet mechanical behaviors are very complex and often described using several metrics including the low-tensile (toe) stiffness, high-tensile (calf) stiffness, or anisotropy index \cite{Meador2020_2}. Consideration of these metrics would likely require a more complex structurally informed constitutive model \cite{Zhang2016} in which the parameters are directly connected to microstructural properties. As such, we chose to simplify the description of the leaflet mechanical behaviors to the general tissue extensibility since it is an overall descriptor of the valve mechanical behavior. However, future extensions of this work could consider how uncertainty or imprecise descriptions of the microstructural properties influence comparisons of valve simulations.

Notably, the valve tissue may adapt to changes in its functional environment, which could in turn change the valve functional environment and cyclically result in long-term progression towards "homeostasis"\cite{Meador2020}. As such, studies into the multiscale properties of valve tissues are evolving, especially within the context of specific valve pathologies~\cite{Thomas2019, rossini2021, Ahmed2021, Poulis2023}. However, currently, relatively little is understood about the interrelationship between valve pathology and tissue mechanical behaviors. Therefore, we assumed the same mechanical properties across valve pathology in our FE models as the disease-driven changes to the valve tissue mechanical behaviors are unknown. We have shown that these disease-driven differences in mechanics (so long as they are within 15\% of a given reference) will not alter the relative ordering of valve pathologies. However, further work is needed regarding the relationship between tissue mechanics, valve pathology, and the durability of valve repair \cite{El-Tallawi2021, Narang2021, marsan2021}. This is an exploratory study seeking to expand the applicability of computational tools to scenarios with no \textit{a priori} knowledge of heart valve properties, but future applications of this open-source computational framework require further experimental validation. 

\section{Conclusion} 
This study has provided the first insight into how tissue extensibility influences FE simulations of atrioventricular valve function. We found that the relative ordering of metrics of valve function and mechanics remained consistent across valve models when varying tissue extensibility so long as the stretches remained within $15\%$ of the reference MV leaflet stretches. As such, FE simulations of atrioventricular valve function may be used to explore how differences and alterations in valve structure affect heart valve function even in populations where material properties are not precisely known.  

\section{Acknowledgment}
This work was funded by the Cora Topolewski Cardiac Research Fund At the Children's Hospital of Philadelphia (CHOP), The Pediatric Valve Center Frontier Program at CHOP, the Additional Ventures Single Ventricle Research Fund, an Additional Ventures Single Ventricle Research Fund Expansion Award, and the National Institutes of Health (NIH R01 HL153166, R01 GM083925, and U24 EB029007). In addition, both WW and DWL were supported by NHLBI T32 HL007915. The authors would also like to acknowledge Professor Akil Narayan from the University of Utah, for providing his expert insights on the statistical approach.

\newpage
\appendix
\section{Additional results for regurgitant MVs with similar geometries} \label{appendix:a}
We provide additional profiles of mechanical and functional metrics for the P2 and annular dilation models. 
\begin{figure}[h!]
\centering
\includegraphics[width=0.9\textwidth]{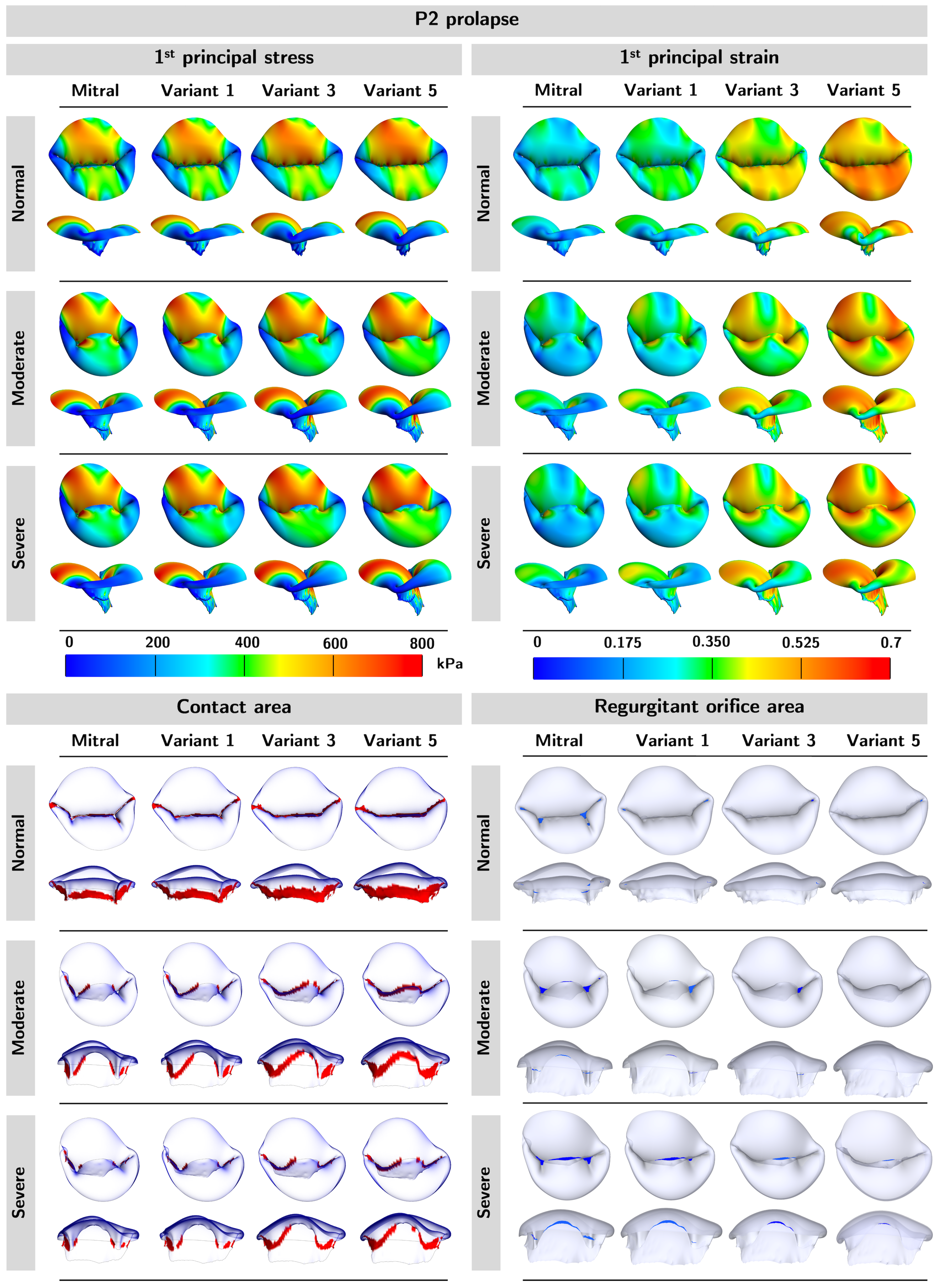}
\caption{\textbf{Profiles of mechanical and functional metrics for P2 prolapse models.} We observed an increase in $1^\text{st}$ principal stress and strain, and contact area but a decrease in ROA as material variants became softer. Additionally, models with higher regurgitation had higher stress, strain, regurgitant area, and lower contact area.}
\end{figure}

\begin{figure}[htbp]
\centering
\includegraphics[width=1\textwidth]{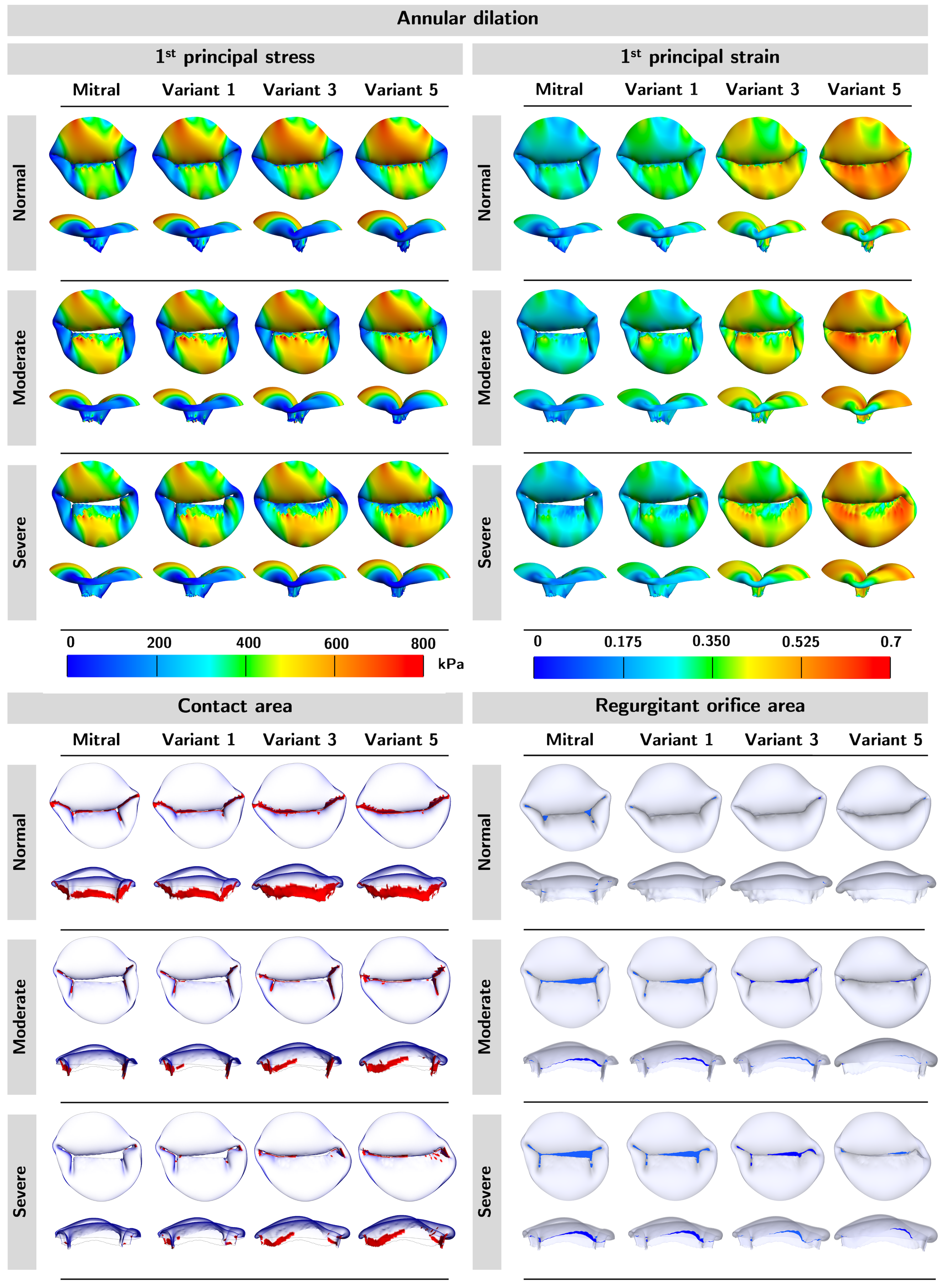}
\caption{\textbf{Mechanical and functional metric profiles for annular dilation models.} We observed an increase in $1^\text{st}$ principal stress, $1^\text{st}$ principal strain, and contact area but a decrease in ROA as the material variants became softer. The models appeared to have higher stresses on the posterior leaflets but lower on the anterior leaflets. The strains were similar across models with varying degrees of regurgitation. Lastly, we observed higher regurgitant and lower contact areas as regurgitation severity increased.}
\end{figure}

\begin{figure}[htbp]
\centering
\includegraphics[width=1\textwidth]{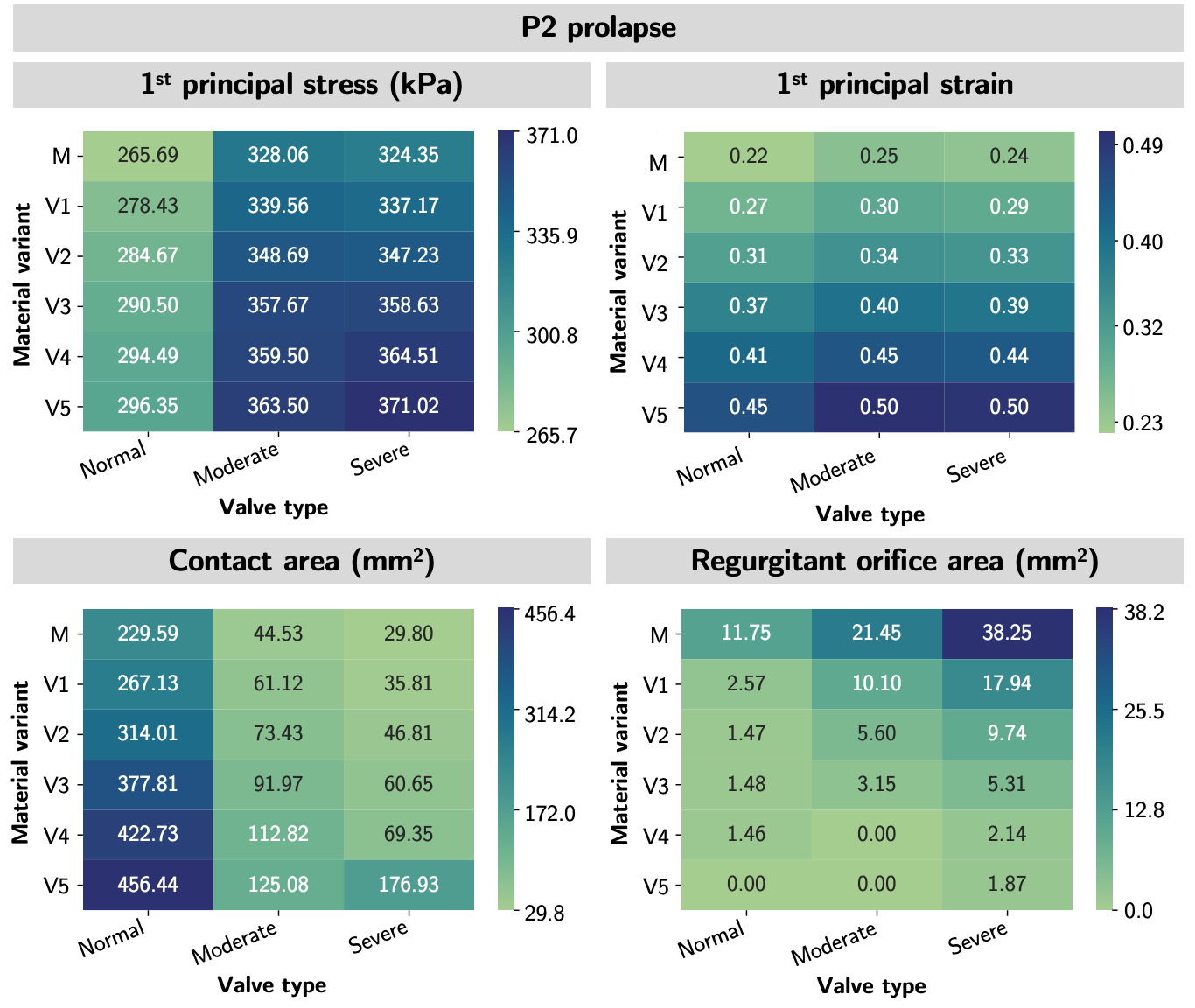}
\caption{\textbf{Mechanical and functional metric profiles for P2 prolapse models.} The heatmaps demonstrated the trend of the mechanical and functional metrics by a) regurgitation severity (across columns) and 2) by material variant (across rows). The relative ordering of the mechanical and functional metrics was mostly preserved in both scenarios. Material variants: M-Mitral; V1-Variant 1; V2-Variant 2; V3-Variant 3; V4-Variant 4; V5-Variant 5. }
\end{figure}

\begin{figure}[htbp]
\centering
\includegraphics[width=1\textwidth]{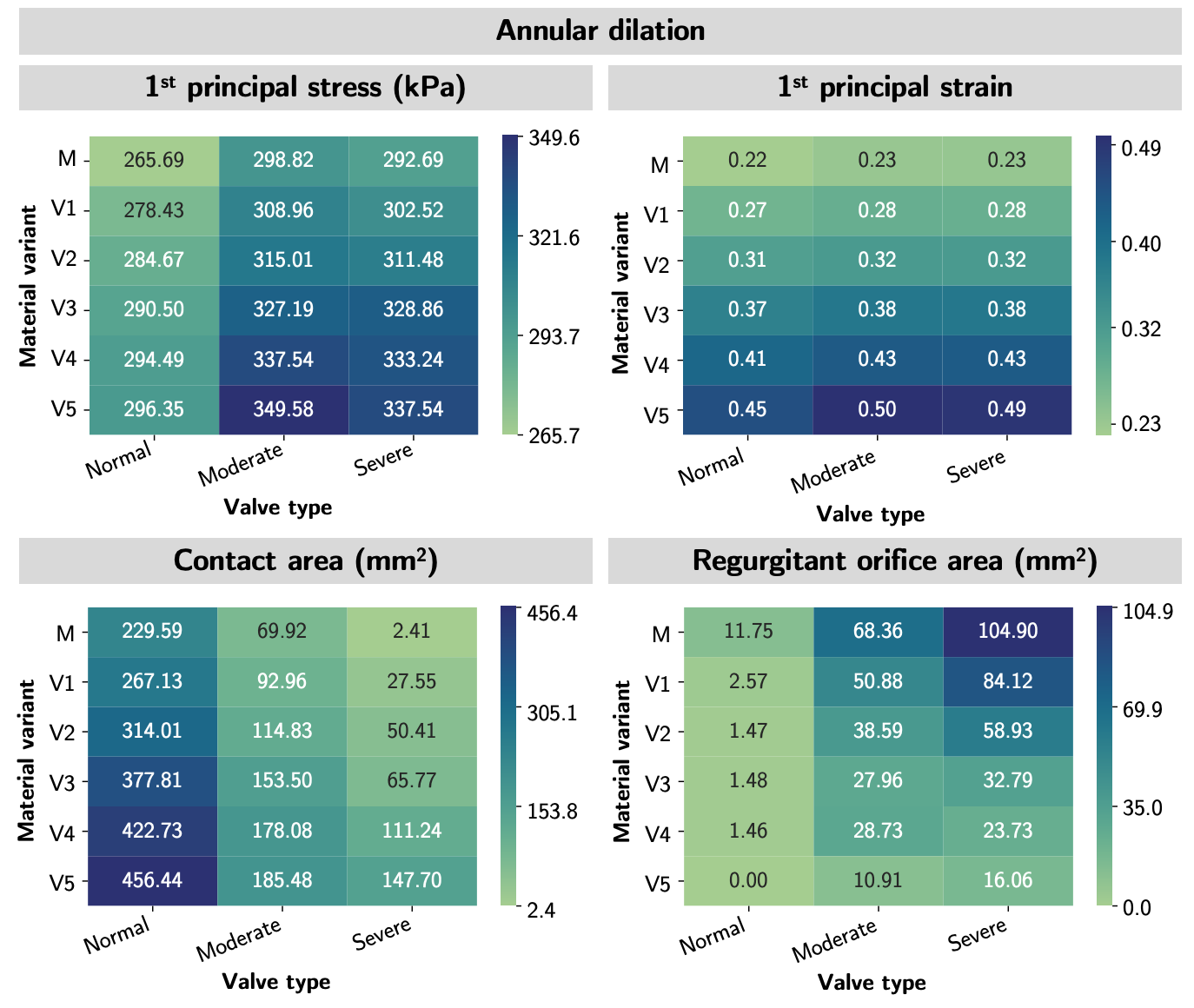}
\caption{\textbf{Mechanical and functional metric profiles for annular dilation models.} The heatmaps demonstrated the trend of the mechanical and functional metrics by a) regurgitation severity (across columns) and 2) by material variant (across rows). The relative ordering of the mechanical and functional metrics was mostly preserved in both scenarios. Material variants: M-Mitral; V1-Variant 1; V2-Variant 2; V3-Variant 3; V4-Variant 4; V5-Variant 5.}
\end{figure}

\newpage
\newpage 
\section{Additional sensitivity analysis results} \label{appendix:b}

\begin{figure}[h!]
\centering
\includegraphics[width=1\textwidth]{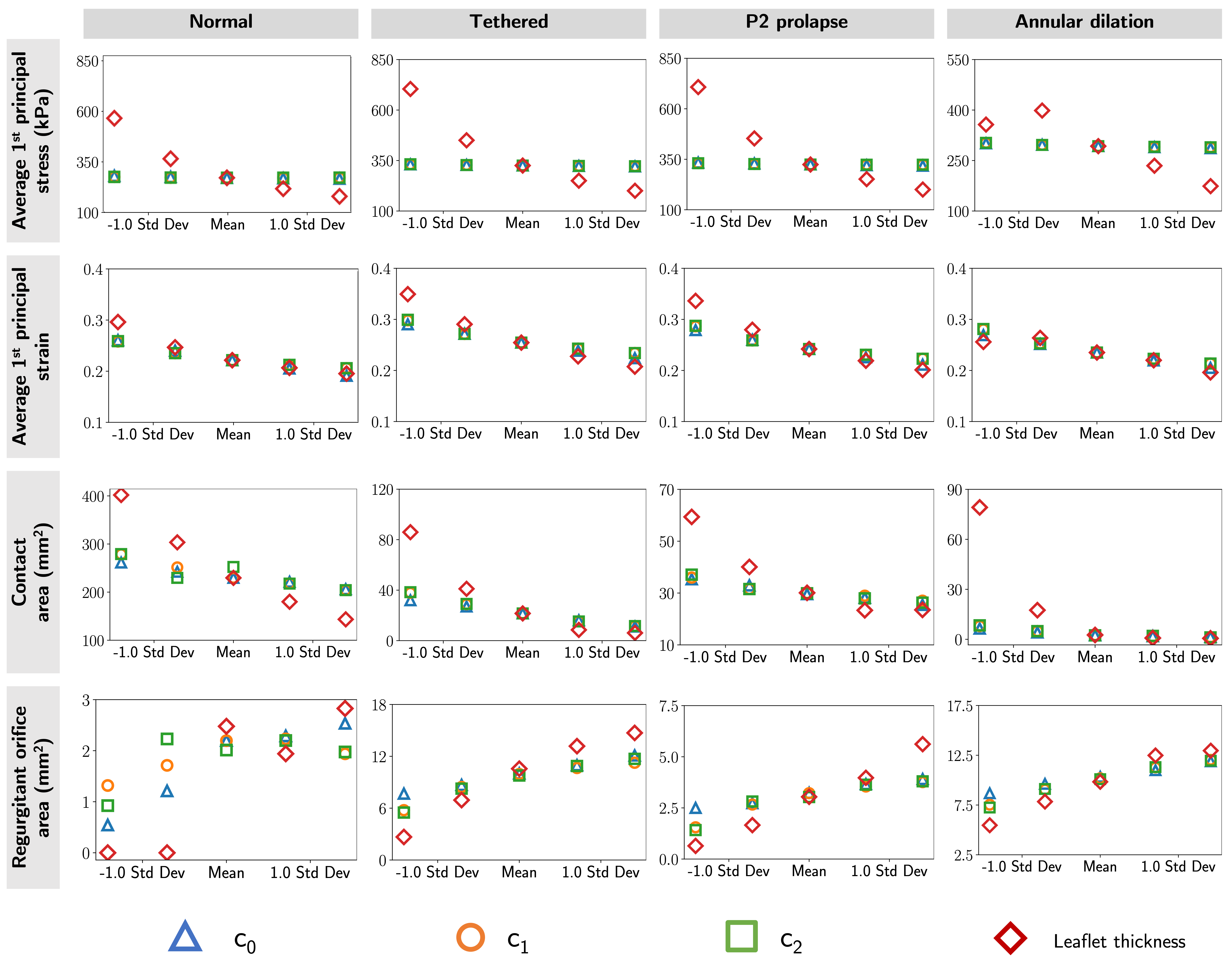}
\caption{\textbf{Sensitivity results.} The material constants appeared to have negligible effects on the average principal stress. Meanwhile, higher leaflet thickness yielded lower leaflet stress. Further, higher leaflet thickness, $c_0$, $c_1$, and $c_2$ led to lower leaflet strain and contact area but higher ROA. } \label{sensitivity::lineplots}
\end{figure}

\bibliographystyle{unsrt}  
\bibliography{library}

\end{document}